\documentclass[journal,article,submit,pdftex,moreauthors]{Definitions/mdpi}

\usepackage{ulem}

\firstpage{1} 
\pubvolume{1}
\usepackage{lineno}
\issuenum{1}
\articlenumber{0}
\pubyear{2023}
\copyrightyear{2023}
\datereceived{ } 
\daterevised{ } 
\dateaccepted{ } 
\datepublished{ } 
\hreflink{https://doi.org/} 

\Title{Exploring the distribution and impact of bosonic dark matter in neutron stars}

\TitleCitation{Exploring the distribution and impact of bosonic dark matter in neutron stars}

\Author{Davood Rafiei Karkevandi $^{1,2,*}$\orcidD{} , Mahboubeh Shahrbaf $^{3,4}$\orcidA{}, Soroush Shakeri $^{1,2}$\orcidS{}, and Stefan Typel $^{5,6}$\orcidC{}}

\AuthorNames{Davood Rafiei Karkevandi, Mahboubeh Shahrbaf, Soroush Shakeri, and Stefan Typel}

\AuthorCitation{Rafiei Karkevandi, D.; Shahrbaf, M.; Shakeri, S.; Typel, S.}

\address{%
$^{1}$ \quad
Department of Physics, Isfahan University of Technology, Isfahan 84156-83111, Iran; \\ \  \ \  \ \  \ \ 
 d.rafiei@alumni.iut.ac.ir; s.shakeri@iut.ac.ir \\
$^{2}$ \quad ICRANet-Isfahan, Isfahan University of Technology, 84156-83111, Iran.\\
$^{3}$ \quad Incubator of Scientific Excellence--Centre for Simulations of Superdense
Fluids, University of Wroclaw; m.shahrbaf46@gmail.com\\
$^{4}$ \quad Frankfurt Institute for Advanced Studies,
Ruth-Moufang-Str. 1, D-60438 Frankfurt am Main, Germany.\\
$^{5}$ \quad Technische Universit\"{a}t Darmstadt, Fachbereich Physik, Institut f\"{u}r Kernphysik,
   Schlossgartenstra\ss{}e 9, D-64289 Darmstadt, Germany; stypel@ikp.tu-darmstadt.de\\
   $^{6}$ \quad GSI Helmholtzzentrum f\"{u}r Schwerionenforschung GmbH,
Theorie, Planckstra\ss{}e 1, D-64291 Darmstadt, Germany.}

\corres{Correspondence: d.rafiei@alumni.iut.ac.ir}

\abstract{The presence of dark matter (DM) within neutron stars (NSs) can be introduced by different accumulation scenarios in which DM and baryonic matter (BM) may interact only through the gravitational force. In this work, we consider asymmetric self-interacting bosonic DM which can reside as a dense core inside the NS or form an extended halo around it. It is seen that depending on the boson mass ($m_{\chi}$), self-coupling constant ($\lambda$) and DM fraction ($F_{\chi}$), the maximum mass, radius and tidal deformability of NSs with DM admixture will be altered significantly. The impact of DM causes some modifications in the observable features induced solely by the BM component. Here, we focus on the widely used nuclear matter equation of state (EoS) called DD2 for describing NS matter. We show that by involving DM in NSs, the corresponding observational parameters will be changed to be consistent with the latest multi-messenger observations of NSs. It is seen that for $m_{\chi}\gtrsim200$ MeV and $\lambda\lesssim2\pi$, DM admixed NSs with $4\%\lesssim F_{\chi}\lesssim20\%$ are consistent with the maximum mass and tidal deformability constraints.
}

\keyword{Bosonic Dark matter, Neutron star, Two-fluid TOV equations, Tidal deformability} 

\begin{document}

\section{Introduction}

Owing to the fact that DM constitutes the majority of matter in galaxies, several noteworthy ideas have been proposed about the presence of DM inside compact astrophysical objects \cite{Baryakhtar:2022hbu,Leane:2020wob,Bramante:2023djs,Leane:2022hkk,Ellis:2018bkr,Nelson:2018xtr,Ryan:2022hku,Chan:2021gcm,Liang:2023nvo}. Among them, NSs, due to their content of high-density matter and extreme gravitational potential, provide an interesting astronomical environment where sizeable amounts of DM may be accumulated based on various scenarios \cite{Karkevandi:2021ygv,Shakeri:2022dwg,Diedrichs:2023trk,Rutherford:2022xeb,Giangrandi:2022wht}. In this regard, the accretion of DM  might occur during star evolution or over the lifetime of a NS. It is argued that high capture rates are more likely to happen towards the center of galaxies, where the density of DM is increasing \cite{PhysRevD.102.063028,Deliyergiyev:2023uer,Bhattacharya:2023stq}. Moreover, DM production inside NSs and supernova explosions can lead to large values of the DM fraction \cite{Shahrbaf:2023uxy,Nelson:2018xtr,Ellis:2018bkr,Shahrbaf:2022upc}. Among various models, a neutron decay anomaly by assuming a dark sector attracts considerable attention \cite{Berryman:2022zic,Shirke:2023ktu,Husain:2022bxl}. It is notable to say that gravitationally stable objects so-called dark stars, may be formed entirely from DM  \cite{Maselli_2017,Pitz:2023ejc,Cassing:2022tnn,Eby:2015hsq}, which can be considered as a capturing center for BM or even merge with NSs resulting in the formation of a mixed compact object \cite{Ellis:2018bkr,Dietrich:2018jov,Clough:2018exo}. 

In recent years, multi-messenger observations of NSs, have provided a unique opportunity to probe their internal structure and the possible existence of exotic configurations including DM \cite{Huth:2021bsp,Raaijmakers:2021uju}. The impact of DM on NS properties has been investigated in various studies proposing a wide range of smoking guns  \cite{Hippert:2022snq,Collier:2022cpr,RafieiKarkevandi:2021hcc,Dengler:2021qcq,Routaray:2022utr,Cronin:2023xzc,Jockel:2023rrm}. Generally, the presence of DM can be considered through two scenarios, i) a single-fluid object for which DM and BM interact also non-gravitationally via an EoS \cite{Shahrbaf:2022upc,Panotopoulos_2017,Das:2020vng,Lourenco:2022fmf,Shahrbaf:2023uxy,Routaray:2023spb,Guha:2024pnn}. ii) Two-fluid DM admixed NSs where DM and BM are described by two EoSs and interact only by the gravitational force \cite{Sandin_2009,Ciarcelluti:2010ji,Karkevandi:2021ygv,Rezaei:2023iif,Thakur:2023aqm,Rezaei_2017,Gleason:2022eeg,Sun:2023cqr}. For the former approach, DM is distributed throughout the whole NSs while for the latter, DM can form a dense core inside NSs or an extended halo around it. Furthermore, in the single-fluid mixed object, DM usually softens the EoS, however, in the case of two-fluid models, DM indicates both softening and stiffening behavior due to the formation of DM core or halo, respectively.  Therefore, considering both approaches, it is seen that asymmetric DM  could affect the maximum mass, radius and tidal deformability of NSs significantly \cite{Karkevandi:2021ygv,Shakeri:2022dwg,Ellis:2018bkr,Nelson:2018xtr,Shahrbaf:2022upc}. Moreover, for asymmetric DM, in which a particle-antiparticle asymmetry exists in the dark
sector \cite{Kaplan:2009ag}, and also self-annihilating (symmetric) DM, the cooling and heating of NSs could be altered which may help to interpret the thermal evolution of NSs \cite{Avila:2023rzj,AngelesPerez-Garcia:2022qzs,Chatterjee:2022dhp,Alvarez:2023fjj,Nguyen:2022zwb}. In addition, DM core or halo formations within NSs can influence the inspiral phase of their binary systems and also the corresponding gravitational wave (GW) signals especially in the post-merger stage \cite{Bauswein:2020kor,Bezares:2019jcb,Ellis:2017jgp,Emma:2022xjs,Ruter:2023uzc,Das:2021wku}. Thanks to the precise mass-radius measurements based on pulse-profile modeling by the NICER telescope \cite{Miller:2019cac,Riley:2021pdl,Watts:2019lbs}, a novel approach has been introduced to probe DM halo around NSs \cite{Miao:2022rqj,Shakeri:2022dwg}, where the variation of the minima in pulse profiles can be severed as a notable signature of DM halo. 

Among various DM candidates, bosonic DM models such as scalar fields, axions and sexaquarks are of great interest in
various aspects of astrophysics and have been examined via their impacts on compact objects \cite{Shakeri:2022usk,Shahrbaf:2022upc,Chavanis:2022fvh}.
In this research in order to model DM, we consider a self-repulsive complex scalar field with sub-GeV bosonic particles and coupling constant in the order of unity. Our DM EoS was first used in the pioneering work by Colpi et al. \cite{Colpi:1986ye} to describe boson stars \cite{Visinelli:2021uve,Liebling:2012fv}, and more recently revisited to model the two-fluid DM admixed NSs \cite{Karkevandi:2021ygv,RafieiKarkevandi:2021hcc,Shakeri:2022dwg}. For the BM component, we employ a widely used and well-known nuclear matter EoS called DD2. Our utilized DD2 EoS is a specific parameterization for the density-dependent meson couplings in a generalized relativistic density functional (GRDF) for nuclear matter which consider $\sigma$, $\omega$, and $\rho$ mesons as exchange particles for describing the effective in-medium interaction of nucleons \cite{Typel:2009sy}. Due to the unknown structure of DM admixed NSs, various nuclear matter EoSs including the ones with phase transition, strange quark matter, and hyperonic degrees of freedom have been proposed to be considered in the mixed compact objects \cite{DelPopolo:2020pzh,Ferreira:2022fjo,Lenzi:2022ypb,Yang:2023haz,Lopes:2023uxi,Lopes:2022efy,Sen:2022pfr,Jimenez:2021nmr}.

 Mass and radius measurements of NSs through observations of radio, optical and X-ray emissions provide reliable bounds for the maximum mass, $M_{max}\gtrsim 2M_{\odot}$ \cite{Miller:2021qha,Romani:2022jhd,NANOGrav:2019jur}, and the radius of $1.4M_\odot$ NS, $R_{1.4}\gtrsim 11$ km \cite{Riley:2019yda,Huth:2021bsp,Dietrich:2020efo}. Furthermore, 
 in addition to constraints for mass and radius, the GW detection of binary NS mergers by the LIGO/Virgo collaboration \cite{Abbott_2017}, has introduced the tidal deformability ($\Lambda$) as a novel parameter to measure the deformation response of compact objects in a binary system  \cite{Hinderer:2009ca}. The tidal deformability,  which is highly sensitive to the
EoS of compact objects and their compactness, is inferred from the detected GW signal in GW170817 event and for a $1.4M_\odot$ NS has an upper limit, $\Lambda_{1.4}\lesssim 580$ \cite{Abbott:2018exr}. These multi-messenger observations have been used extensively to constrain dense nuclear matter properties and to explore the possible existence of DM inside NSs by constraining the DM parameter space such as particle mass, coupling constant and its fraction inside the mixed object \cite{Karkevandi:2021ygv,Shakeri:2022dwg,Giangrandi:2022wht,Rutherford:2022xeb,Thakur:2023aqm,Thakur:2024mxs}. Recently, by investigating the consistency of the DM admixed NS observable features with the above-mentioned astrophysical limits, an excluded region for DM parameters and a maximum possible fraction of DM have been obtained in \cite{Karkevandi:2021ygv,Shakeri:2022dwg}.

DD2 EoS is considered one of the preeminent nuclear EoSs for which the density dependence 
of the couplings is adjusted to describe the properties of atomic nuclei 
around saturation density. However, since DD2 is a stiff EoS, it does not yield a value for $\Lambda_{1.4}$ that is consistent with astrophysical constraints. 
Considering the aim of the paper, we investigate the impact of bosonic DM on the maximum mass, radius, and particularly the tidal deformability derived from the DD2 EoS. Indeed, we are going to
show how the presence of DM may modify the observable characteristics of NSs derived from nuclear EoSs. The allowed bosonic DM parameter space and fraction will be investigated for which all of the aforementioned parameters are in agreement with observational bounds. Our result indicates a modification of NS features obtained from the considered nuclear matter model thanks to the presence of DM. In light of current and upcoming promising astrophysical missions, we provide strong hints about the possibility of the existence of DM in NSs and its influence on various measurable parameters. 

The paper is organized as follows. In Sec. \ref{sec2} we present a two-fluid formalism for DM admixed NSs in addition to introducing the BM and DM EoSs. Various distributions of bosonic DM in NSs are considered in Sec. \ref{sec3}. The impact of DM on the tidal deformability of the mixed objects is examined in Sec. \ref{sec4}, in Sec. \ref{sec5} the parameter space of DM model is considered in light of the latest multi-messenger constraints and finally the conclusion is given in \ref{sec6}. In this paper, we use units in which $\hbar=c=G=1$.

\section{Two-fluid DM admixed NSs}
\label{sec2}

Owing to the fact that in two-fluid DM admixed NSs, BM and DM interact only via the gravitational force, the energy-momentum tensors are conserved separately $T^{\mu \nu}=T^{\mu \nu}_{DM}+T^{\mu \nu}_{BM}$. Thus, solving the Einstein equation leads to the two-fluid Tolman-Oppenheimer-Volkof (TOV) relations as follows \cite{Sandin_2009,Ciarcelluti:2010ji,Xiang:2013xwa}
\begin{eqnarray}\label{e6}
\frac{dp_{\text{B}}}{dr}&=& -\left( p_{\text{B}} +\epsilon_{\text{B}} \right) \frac{M+4\pi r^{3}p}{r(r-2M)}\,,\\ 
\frac{dp_{\text{D}}}{dr}&=& -\left( p_{\text{D}} +\epsilon_{\text{D}} \right) \frac{M+4\pi r^{3}p}{r(r-2M)}, \label{e6d}
\end{eqnarray}
Here $p_{B}$, $p_{D}$, $\epsilon_{B}$ and $\epsilon_{D}$ are the corresponding pressure and energy density for each of the fluids, respectively, 
where the B and D indices stand for
BM and DM components. Moreover, $p=p_{B}+p_{D}$ denotes the summation of pressure for both BM and DM and $M=\int_{0}^{r} 4\pi r^{2} \epsilon_B (r) dr + \int_{0}^{r} 4\pi r^{2} \epsilon_D (r) dr$ is the mass inside a radius $r$. Depending on the point where the pressure of one of the fluids gets zero first in the numerical calculations of the two-fluid TOV equations, mainly two possible DM distributions can be found within NSs, called DM core and DM halo configurations. For both cases, the total mass of the object is $M_{T}=m_{B}(R_{B})+m_{D}(R_{D})$ where $m_{B}(R_{B})$ and $m_{D}(R_{D})$ are the enclosed masses of BM and DM, respectively. If 
 DM resides as core inside NS,  $R_{B}>R_{D}$, and the outermost radius of the object is the BM radius. However, when DM forms a halo around a NS, $R_{D}>R_{B}$, and the radius of the object is defined by the DM fluid. It is worth mentioning that for both possible DM configurations, the visible radius of the star is $R_{B}$. Moreover, the amount of DM inside the mixed object which is considered a model parameter in our analysis is defined as $F_{\chi}=\frac{m_{D}{(R_{D})}}{M_{T}}$.

In this research, the DM fluid is described with the following Lagrangian
\begin{eqnarray}
\mathcal{L}=\frac{1}{2}\partial_\mu\phi^*\partial^\mu\phi
-\frac{m_\chi^2}{2}\phi^*\phi-\frac{\lambda}{4}(\phi^*\phi)^2,
\end{eqnarray}
where a self-interacting complex scalar field, $\phi$, has been considered as the bosonic DM with a repulsive interaction defined by  $V(\phi)=\frac{\lambda}{4}|\phi|^4$ potential for which $\lambda$ is the dimensionless coupling constant and $m_{\chi}$ is the particle mass \cite{Colpi:1986ye}. In the strong coupling regime, the system can be approximated as a perfect fluid
and the anisotropy of pressure will be ignored, so one can reach the EoS of the self-repulsive bosonic DM as follows 
\begin{equation}\label{e1}
P=\frac{m_{\chi}^{4}}{9\lambda}\left( \sqrt{1+\frac{3\lambda}{m_{\chi}^{4}}\rho}-1\right)^{2}.
\end{equation}
Where $P$ and $\rho$ are the pressure and density of the self-interacting bosonic DM (see the appendix of \cite{Karkevandi:2021ygv} for more details about the bosonic DM EoS in which a derivation has been proposed based on the mean-field approximation in a flat space-time). Here, the well-known DD2 model is utilized to describe the EoS of BM. The DD2 parameterization was obtained by fitting the properties of finite nuclei such as binding energy, charge and diffraction radii, surface thickness, and spin-orbit splitting \cite{Typel:2009sy}. The predicted nuclear matter parameters at saturation point are consistent with recent constraints. In particular, a saturation density of 0.149065 $\mathrm{fm}^{-3}$, a binding energy per nucleon of 16.02 MeV, and an incompressibility of 242.7 MeV are obtained for the symmetric nuclear matter within DD2 model \cite{Oertel:2016bki}. The symmetry energy at saturation and its slope parameter are given by $J=31.7$ MeV and $L=55.0$ MeV. 
Further, the observable properties of NSs obtained from the DD2 EoS are  $M_{max}=2.4M_{\odot}$, $R_{1.4}=13.15$ km and $\Lambda_{1.4}=681.61$. The large maximum mass, radius and tidal deformability indicate that  DD2 EoS is rather stiff. Even though $M_{max}$ and $R_{1.4}$ are consistent with astrophysical observations, $\Lambda_{1.4}$ is larger than the observational constraint i.e. $\Lambda_{1.4}\lesssim580$ from GW170817. In the following sections, we are going to check how the presence of DM could alter the observable features and modify the tidal deformability. 

\section{DM distributions in NSs}
\label{sec3}

\begin{figure}[h!]
   \centering
	\includegraphics[width=0.49\textwidth]{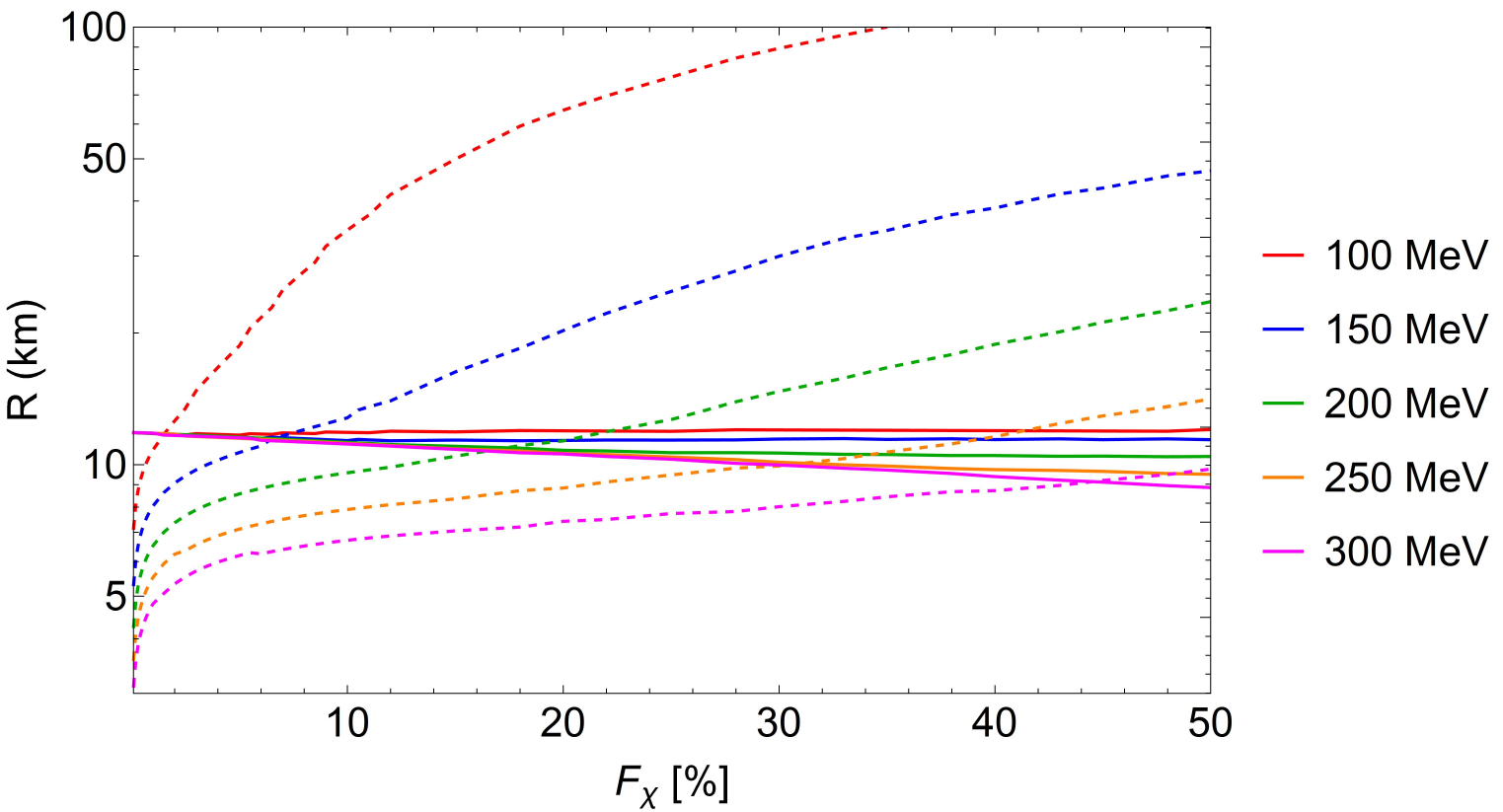}
  \includegraphics[width=0.49\textwidth]{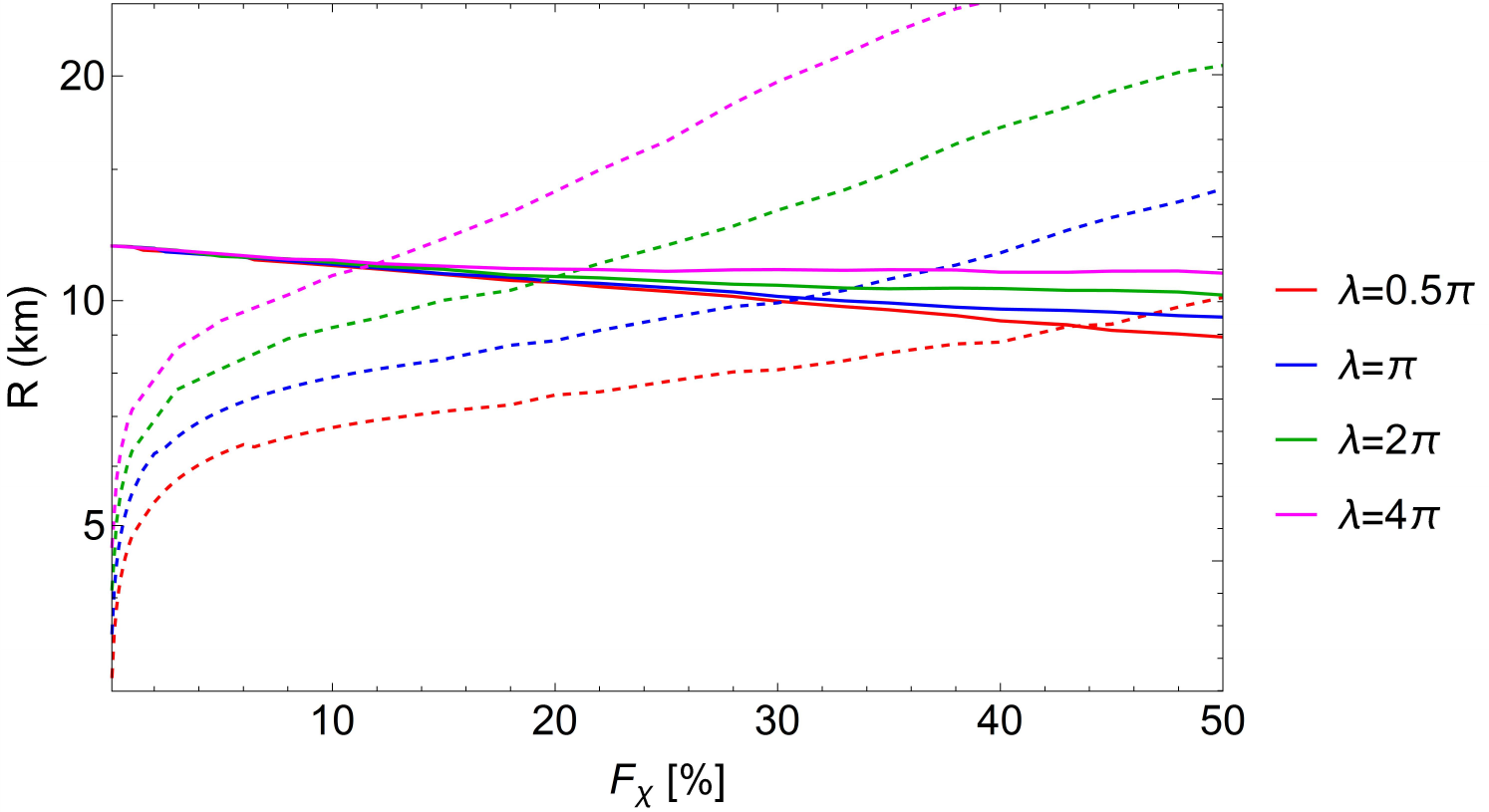}
	\caption{Variation for the radii of BM ($R_{B}$) and DM ($R_{D}$) are shown with respect to the DM fraction ($F_{\chi}$) by solid and dashed lines, respectively. All the radii are related to the maximum mass of DM admixed NSs for various DM model parameters. Note that each colored solid and dashed line corresponds to a specific boson mass (left) or coupling constant (right), as indicated in the legends.  In the left panel, different bosonic particle masses are considered as labeled for $\lambda=\pi$, while the right panel is for coupling constants varied from $0.5\pi$ to $4\pi$ and $m_{\chi}=250$ MeV.  
		\label{FX-R}
	}
	\end{figure}

In this section, we are going to investigate the distribution of self-repulsive bosonic DM inside NSs by considering the DD2 EoS as the BM component. Therefore, in Fig. (\ref{FX-R}) the variation of $R_{B}$ (solid line) and $R_{D}$ (dashed line) is depicted as a function of the DM fraction ($F_{\chi}$) for various boson masses ($m_{\chi}$) and coupling constants ($\lambda$). It should be mentioned that each colored line, both solid and dashed, is associated with a specific value of $m_{\chi}$ (left panel) or $\lambda$ (right panel), as labeled. Note that the radii for each given DM model parameter and fraction correspond to the maximum mass of the DM admixed NSs. For both plots, it is seen that in low $F_{\chi}$ the DM radius (dashed lines) is smaller than the BM  radius (solid lines) and a DM core is formed inside the NS. However, toward higher DM fractions, $R_{D}$ will increase and surpass the corresponding $R_{B}$ leading to a DM halo formation around the BM shell. Therefore, a transition from DM core to DM halo formation can be seen by enhancing $F_{\chi}$ for each considered case.  In the left panel, it is seen that heavier bosons ($m_{\chi}\gtrsim200$ MeV), in comparison to lighter ones, reside mainly in the core inside NSs while for high DM fractions, they produce a small halo around the NS. Moreover, in the right panel where the impact of the coupling constant is considered for a fixed boson mass ($m_{\chi}=250$ MeV), higher couplings result in larger DM halos for a given $F_{\chi}$. Obviously, the threshold DM fraction for which a DM core to DM halo transition occurs will increase by decreasing the coupling constant. Regarding the visible radius of the object ($R_{B}$), both of the plots show a decrease along higher DM fractions which for heavy bosons and lower couplings the rate of reduction is larger.

\begin{figure}[h!]
   \centering
	\includegraphics[width=0.6\textwidth]{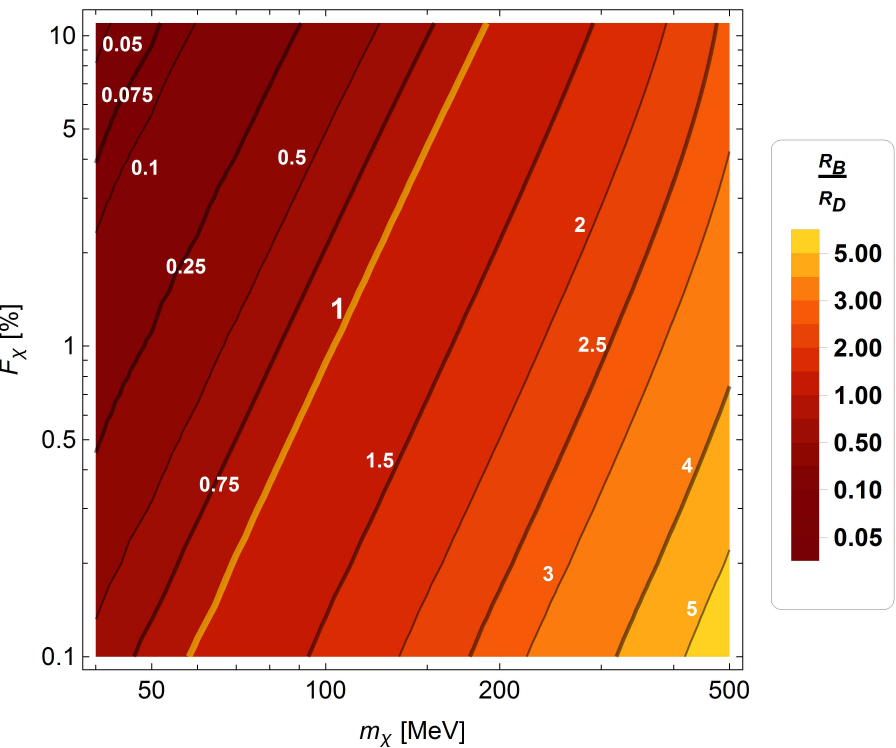}
	\caption{ The contour plot shows $R_{B}/R_{D}$ ratio for a scan over the $F_{\chi}-m_{\chi}$ parameter space considering a fixed coupling constant $\lambda=0.5\pi$ for $1.4M_{\odot}$ DM admixed NSs. The yellow line indicates where $R_{B}=R_{D}$ and the DM-core to DM-halo transition occurs.  
	}
 \label{Rscan}
	\end{figure}

For the sake of completeness, we performed a scan over the $F_{\chi}-m_{\chi}$ parameter space for DM admixed NSs with $M_{T}=1.4M_{\odot}$ and fixed $\lambda=0.5\pi$. The results are presented in Fig. (\ref{Rscan}), which shows the contours representing the values of $R_{B}/R_{D}$. It is seen that light bosons form an extended halo around NSs whilst heavier bosons make a dense DM core inside the star. The yellow line  $R_{B}=R_{D}=1$ depicts the region where DM is distributed inside the whole NS and a transition occurs from DM core to DM halo for any given parameter. Furthermore, for the regions where a halo is formed, higher values of $F_{\chi}$ lead to a very large halo ($R_{B}/R_{D}\lesssim0.1$), however, for the part where a core is formed, lower values of DM fractions result in a very small DM core ($R_{B}/R_{D}\gtrsim3$) in the BM shell. 

Regarding the aforementioned results about DM core and DM halo formations, in the following sections, we are going to investigate the impacts of self-interacting bosonic DM on NSs observable features.   

\section{Tidal deformability of DM admixed NSs}
\label{sec4}
The GW detection of a NS binary merger in the GW170817 event and measuring its tidal deformability provide a novel approach to probe high-density matter and the possibility to explore the existence of more exotic compact objects including DM. In this section, we investigate the tidal deformability ($\Lambda$) of DM admixed NSs for various bosonic DM model parameters and compare it with the pure NS which is modeled by the DD2 EoS. The dimensionless tidal deformability is defined as $\Lambda=\frac23 k_2 \left(\frac{R}{M}\right)^5\,$ in which R and M are the outermost radius and total mass of the object, and $k_{2}$ is the
tidal love number \cite{Hinderer:2007mb} calculated from the two-fluid TOV formalism \cite{Karkevandi:2021ygv}. Notice that $\Lambda$ for  DM admixed NSs with $1.4M_{\odot}$ is highly dependent on the outermost radius of the object which will be varied when DM core or DM halo is formed inside or around NSs.

\begin{figure}[h!]
   \centering
	\includegraphics[width=0.6\textwidth]{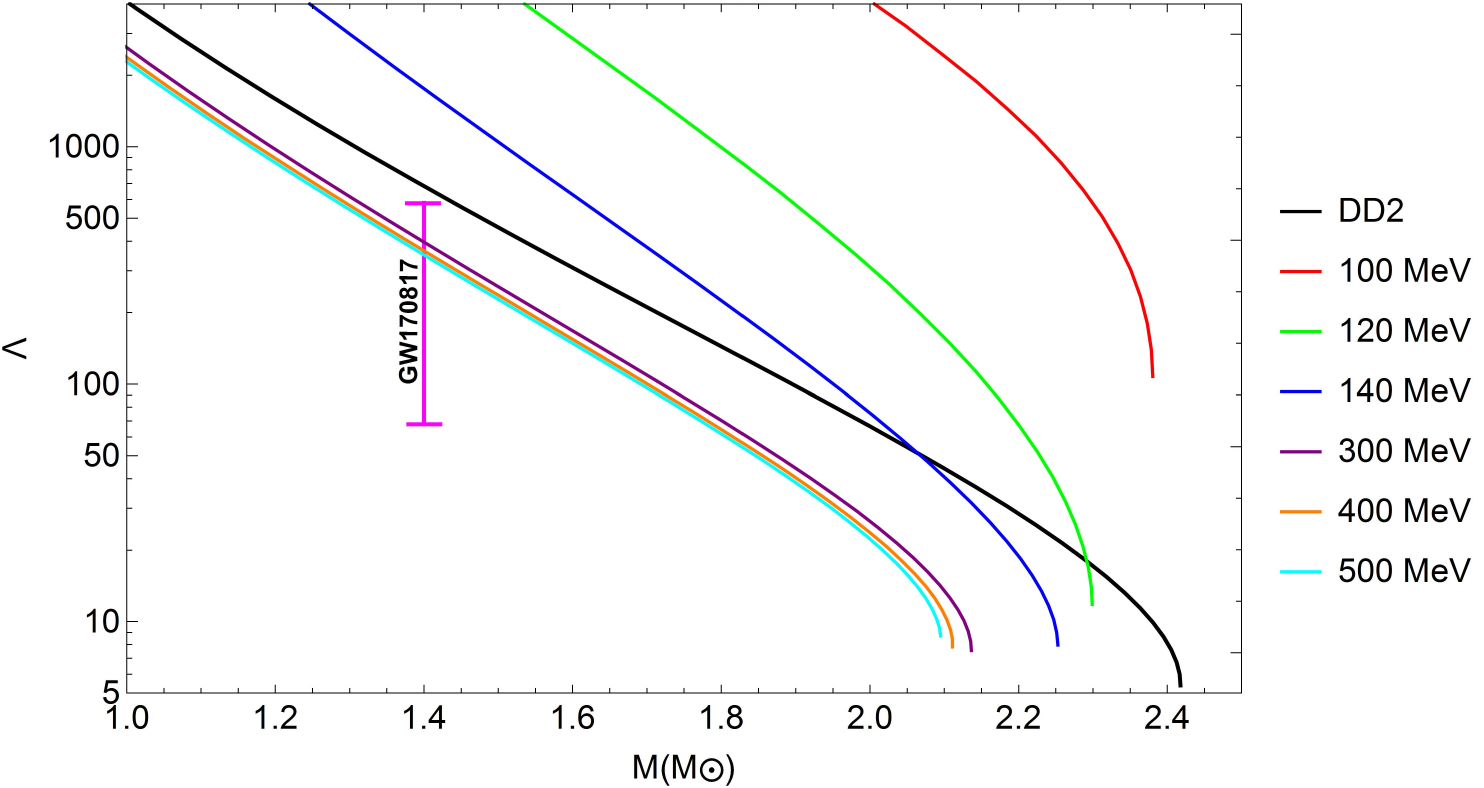}
	\caption{The dimensionless tidal deformability, denoted as $\Lambda$, is presented as a function of the stellar mass for different bosonic particle masses as labeled, considering fixed $F_{\chi}=10\%$ and $\lambda=\pi$. The solid black line represents the $\Lambda-M$ graph for a pure NS without DM. The magenta vertical line signifies the $\Lambda_{1.4}$ constraint derived from the low-spin prior analysis of GW170817, as reported in \cite{Abbott:2018exr}.
		\label{Lam-M1}
	}
	\end{figure}

\begin{figure}[h!]
   \centering
	\includegraphics[width=0.6\textwidth]{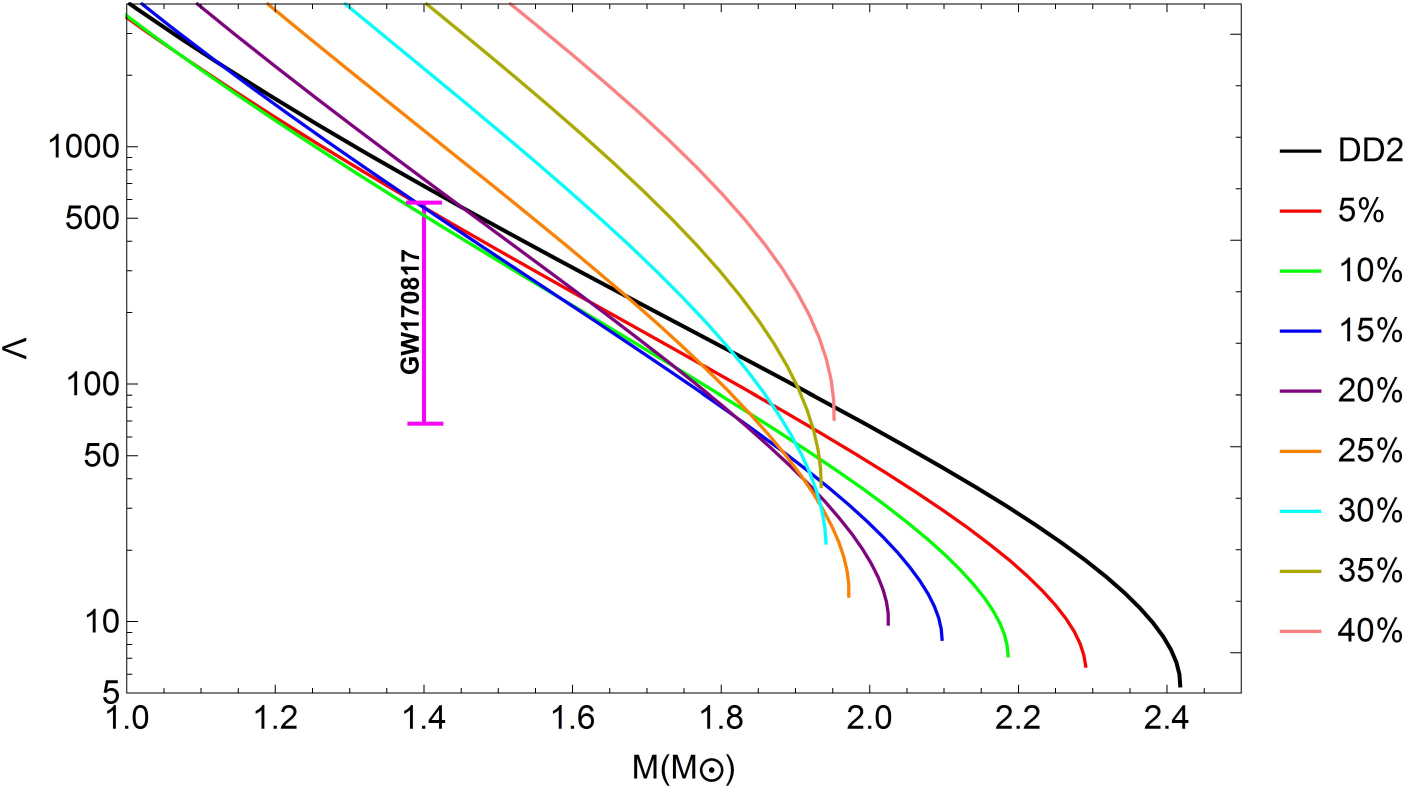}
	\caption{ Similar to Fig.(\ref{Lam-M1}), but for different DM fractions and $m_{\chi}=200$ MeV and $\lambda=\pi$.
		\label{Lam-M2}
	}
	\end{figure}

\begin{figure}[h!]
   \centering
	\includegraphics[width=0.6\textwidth]{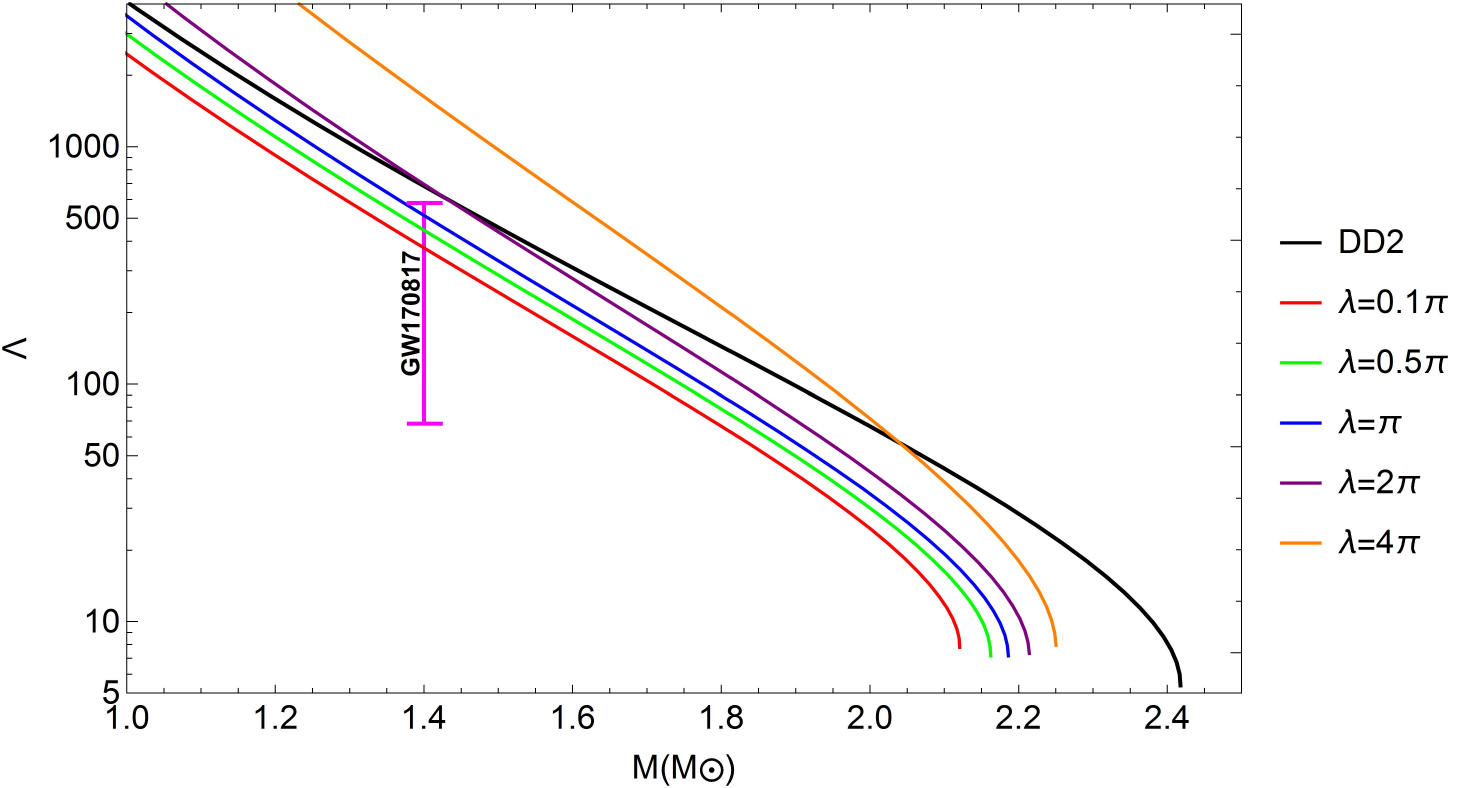}
	\caption{ Similar to Fig.(\ref{Lam-M1}), but for different coupling constants and $m_{\chi}=200$ MeV and $F_{\chi}=10\%$.
		\label{Lam-M3}
	}
	\end{figure}

In the following, Figs. \ref{Lam-M1}, \ref{Lam-M2} and \ref{Lam-M3} indicate the tidal deformability of DM admixed NSs with respect to the mass of the object for various bosonic DM model parameters and fractions. In all of the plots, the curve related to pure BM NS (without DM) is shown by a black solid line and the magenta vertical line depicts the allowed observational range for the tidal deformability of a $1.4M_{\odot}$ NS, $\Lambda_{1.4}=190^{+390}_{-120}$ reported by \cite{Abbott:2018exr}. Obviously, it is seen that the corresponding $\Lambda_{1.4}$ of DD2 EoS is higher than the upper allowed limit for tidal deformability, thus we will examine how bosonic DM will alter this parameter which may become compatible with the GW170817 constraint.

In Fig. (\ref{Lam-M1}), different $m_{\chi}$ are considered for fixed $F_{\chi}=10\%$ and $\lambda=\pi$. We see that light bosons increase $\Lambda$ because of the rise in the outermost radius of the object and the formation of a large DM halo. However, for massive bosons ($m_{\chi}\geq300$ MeV) which make the star more compact, the tidal deformability will be reduced and is consistent with the allowed astrophysical limit.

 Taking various DM fractions into account for  $m_{\chi}=200$ MeV and $\lambda=\pi$, Fig. (\ref{Lam-M2}) illustrates that for $F_{\chi}\lesssim20\%$, the tidal deformability is improved, however, for higher fractions both maximum mass and $\Lambda$ are not consistent with the observational constraints. Indeed, it is displayed that by raising $F_{\chi}$, the tidal deformability is enhanced which is due to a transition from a DM core to a DM halo formation by varying the amount of DM. Moreover in Fig. (\ref{Lam-M3}) we probe the impact of a changing coupling constant on the tidal deformability of DM admixed NSs for $F_{\chi}=10\%$ and $m_{\chi}=200$ MeV. It is seen that by decreasing the coupling constant, the tidal deformability will be reduced and becomes consistent with the observational range while the maximum mass is also in agreement with the $2M_\odot$ limit. However, for $\lambda\gtrsim2\pi$, we see an enhancement in the tidal deformability which is higher than the astrophysical bound because of the DM halo surrounding the NSs.

Therefore, concerning the aforementioned results, in the next section, we will probe the bosonic DM parameter space for which both  $M_{T_{max}}$ and $\Lambda_{1.4}$ are in agreement with the astrophysical limits and indicate the capability of bosonic DM in modifying the observable features related to the BM EoS. 
 
\section{Probing bosonic DM parameter space}
\label{sec5}
In this section, we will examine the bosonic DM model parameters i.e. boson mass and coupling constant and its fraction, by considering observable quantities of the mixed compact object. The total maximum mass and tidal deformability of a $1.4M_{\odot}$ DM admixed NS are taken into account to probe the DM parameter space which is consistent with both the $M_{max}\geq2M_{\odot}$ and  $\Lambda_{1.4}\lesssim580$ constraints. It should be mentioned that DD2 is a stiff EoS which results in a large radius around $1.4M_{\odot}$, $R_{1.4}=13.15$ km, thus the $R_{B_{1.4}}\geq11$km constraint will be generally satisfied for this investigation, however, for more details about the variation of dark and visible radius of the star see \cite{Shakeri:2022dwg}.

\begin{figure}[h!]
   \centering
	\includegraphics[width=0.49\textwidth]{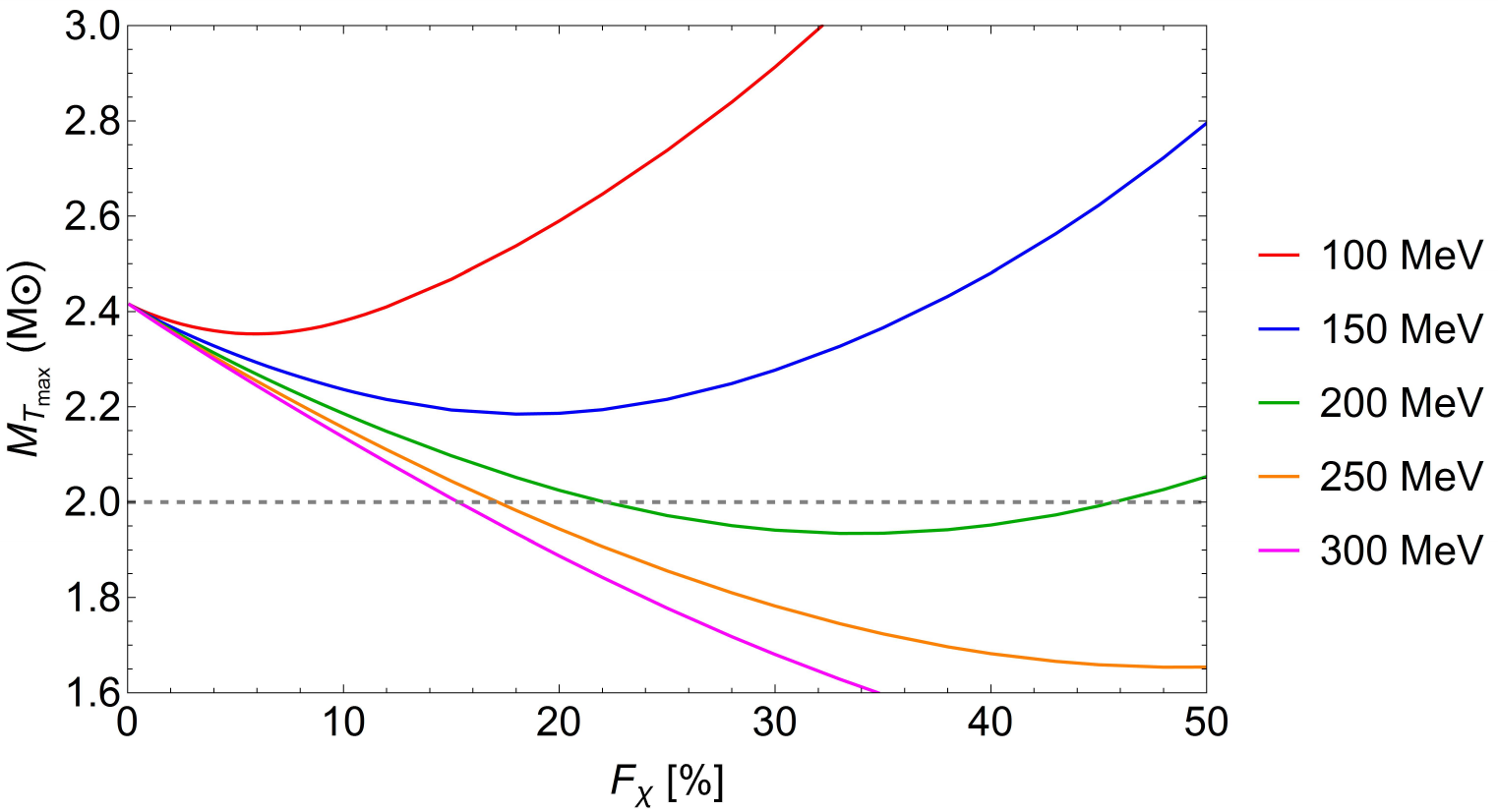}
  \includegraphics[width=0.49\textwidth]{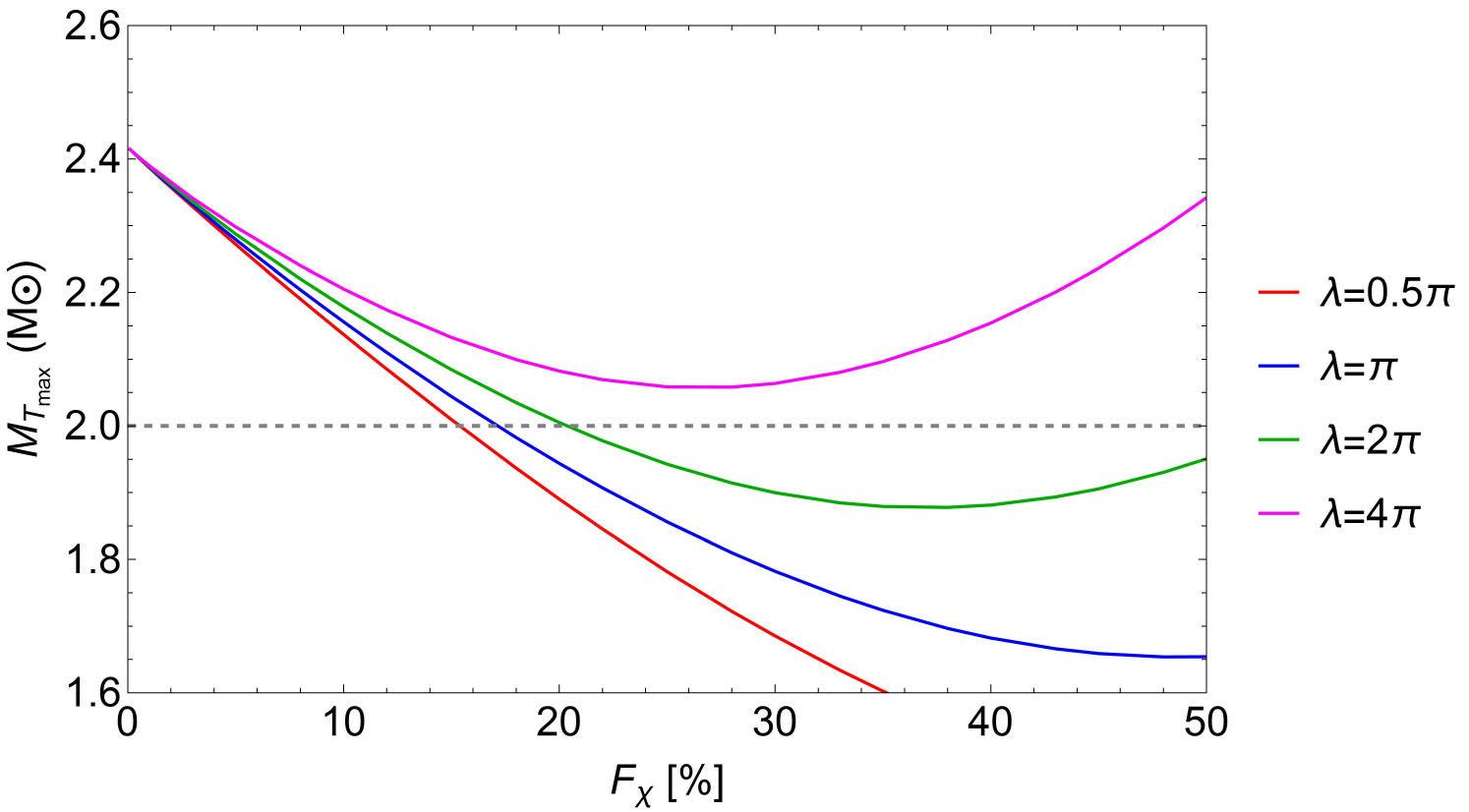}
	\caption{ The change of total maximum mass of DM admixed NSs as a function of $F_{\chi}$ are indicated for different DM model parameters. The left panel corresponds to various boson masses and $\lambda=\pi$, while the right panel is related to several coupling constants as labeled and $m_{\chi}=250$ MeV. The gray dashed line depicts the $2M_{\odot}$ limit for the mass of NSs.
  \label{MT-FX}
	}
	\end{figure}

 Concerning DM core and DM halo formation, Fig. (\ref{MT-FX}) illustrates the variation of $M_{T_{max}}$ with respect to DM fraction.  In both plots, for all of the considered DM parameters, the maximum mass of the object will decrease in low fractions which correspond to the DM core formation. However, by increasing $F_{\chi}$, the maximum mass of the star gradually rises due to the DM halo configuration. In the left panel, it is seen that for light bosons $M_{T_{max}}$ increases sharply and is higher than the $2M_{\odot}$ limit (shown by the gray dashed line) for the whole range of DM fractions. For more massive bosonic particles ($m_{\chi}\gtrsim200$ MeV) there is a DM fraction limit where the maximum mass does not agree with the astrophysical bound and falls below $2M_{\odot}$. Regarding the right panel, it is indicated that for $\lambda=4\pi$, $M_{T_{max}}$ is always consistent with the observational constraint, however, for lower coupling constants,  one can find a DM fraction beyond which $M_{T_{max}}<2M_{\odot}$. Furthermore, both plots demonstrate that for small 
amounts of DM fractions, the total maximum mass of DM admixed NSs is in agreement with  $M_{max}\geq2 M_{\odot}$ constraint for all of the considered boson masses and coupling constants.

\begin{figure}[h!]
   \centering
	\includegraphics[width=0.49\textwidth]{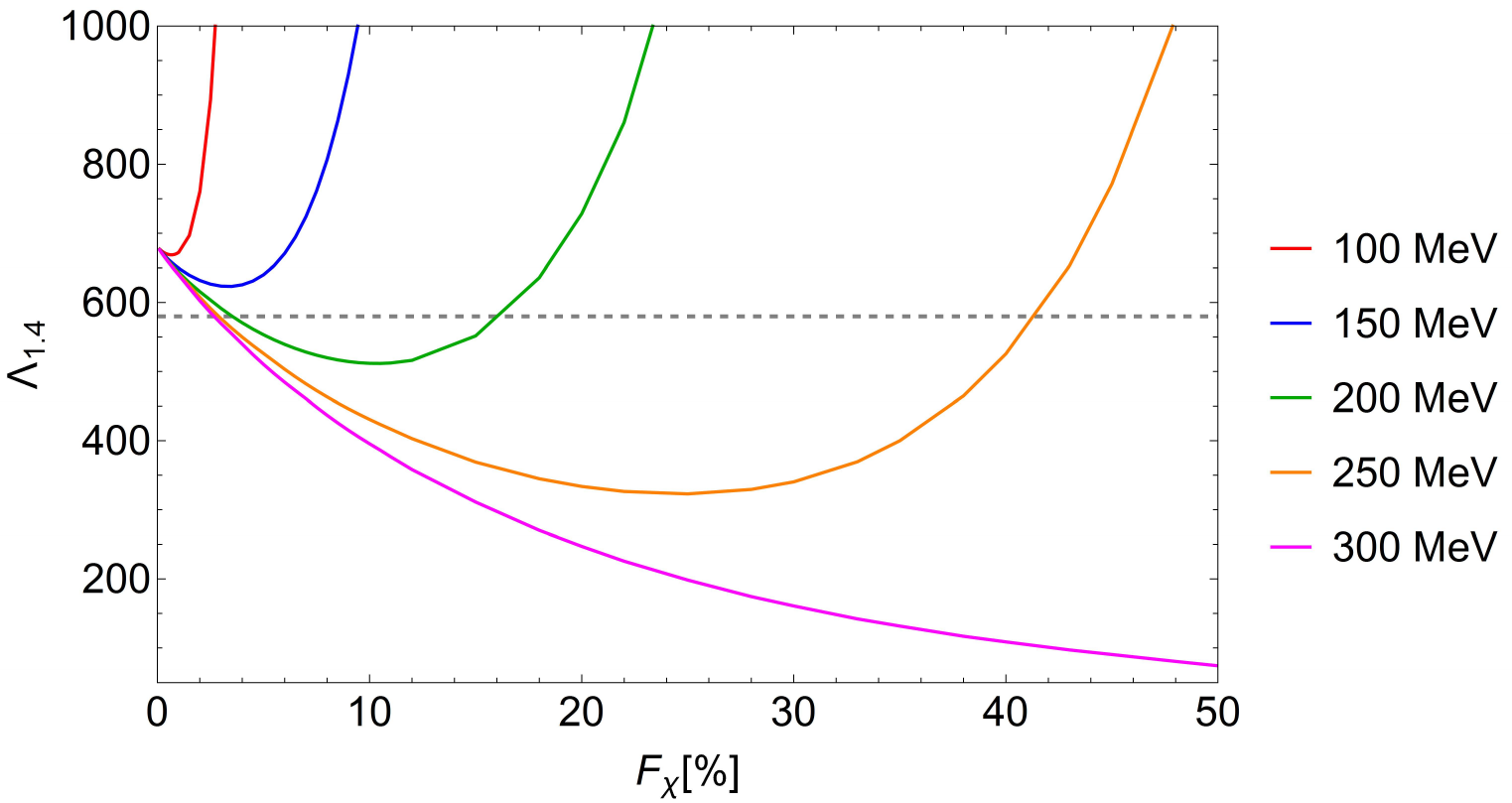}
  \includegraphics[width=0.49\textwidth]{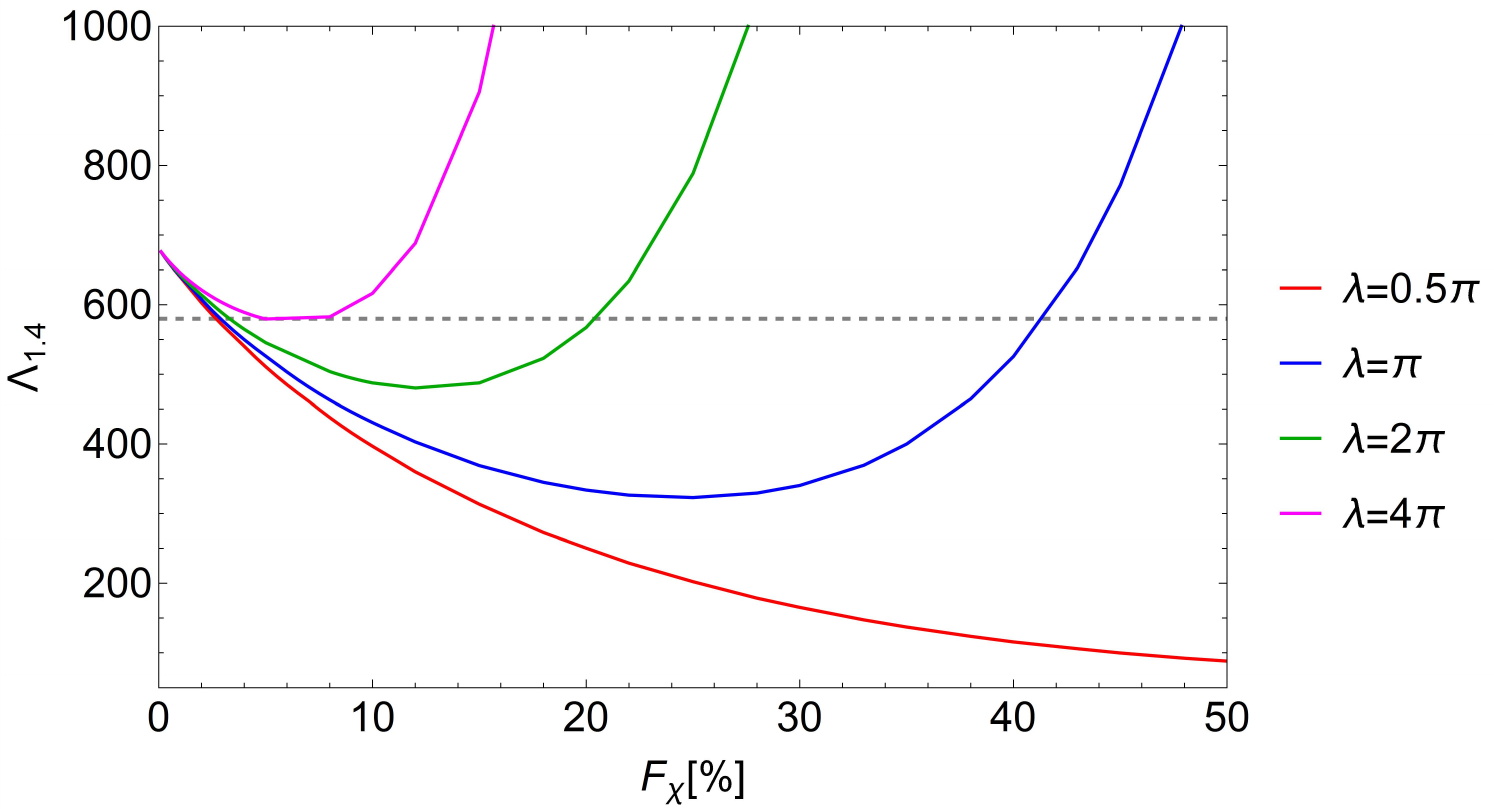}
	\caption{Variation of $\Lambda_{1.4}$ is shown in terms of DM fraction for different boson masses and coupling constants. In the left panel $\lambda=\pi$ is fixed while in the right panel $m_{\chi}=250$ MeV for all of the cases. The maximum observational limit for the tidal deformability of $1.4M_{\odot}$ NS (580) is indicated by the gray dashed line.
  \label{Lam-FX}
	}
	\end{figure}

Considering the fact that $\Lambda_{1.4}$ of DD2 EoS is not consistent with the observational constraint,  Fig. (\ref{Lam-FX}) depicts the change of tidal deformability for $1.4M_{\odot}$ DM admixed NSs as a function of $F_{\chi}$. Various boson masses and a fixed coupling constant $\lambda=\pi$ are considered in the left panel, however, for the right panel $\lambda$ is varied for the boson mass $m_{\chi}=250$ MeV. Generally, in both of the plots, there are some regions where $\Lambda_{1.4}$ is modified by the bosonic DM and becomes less than the gray dashed line which shows the maximum confirmed observational value ($\Lambda_{1.4}=$580). In the left panel, it is seen that for the light bosons, $\Lambda_{1.4}$ is higher than the astrophysical constraint for the whole considered DM fractions which is due to the DM halo formation. For $m_{\chi}\geq200$ MeV there are ranges of the DM fraction for which the $\Lambda_{1.4}\lesssim$580 bound is satisfied thanks to the reduction in the outermost radius of the star by the DM core configuration. Moreover, in the right panel which shows the impact of the coupling constant, it is seen that lower values of $\lambda$ are more compatible with the tidal deformability constraint for a wider range of DM fractions inside NSs. This behavior is caused by the fact that for low coupling constants DM mainly resides as a core which decreases the tidal deformability. Regarding both plots, it is seen that in low DM fractions ($F_{\chi}\lesssim4\%$), $\Lambda_{1.4}$ is higher than 580 for all of the applied DM parameters, while it will be decreased by increasing $F_{\chi}$ and become less than 580 for massive bosons and low coupling constants.

In both Figures \ref{MT-FX} and \ref{Lam-FX}, we observe a decrease and increase in the values of the total maximum mass and tidal deformability by changing the
DM fraction, this behavior is associated with the formation of a DM core or DM halo, respectively. Comparing the variation of $M_{T_{max}}$ in Fig. (\ref{MT-FX}) with the corresponding radius in Fig. (\ref{FX-R}), we note that for  $F_{\chi}<10\%$ where $M_{T_{max}}$ decreases, the radius of the DM fluid is notably smaller than the BM fluid radius. Conversely, for higher  $F_{\chi}$ where $M_{T_{max}}$ increases, the corresponding DM radius is noticeably larger than the BM one. Moreover, changes in the outermost radius due to the DM core and halo formation (see Fig. \ref{Rscan}), significantly affect $\Lambda_{1.4}$ and cause the variation seen in Fig. (\ref{Lam-FX}).

In general, the softness/stiffness of DM EoS is varied with DM model parameters, while the inclusion of DM in NSs leads to the softening of the EoS of a DM admixed NS which results in a decrease of the maximum mass and tidal deformability compared to a pure NS. However, towards higher DM fractions a transition occurs from a softer EoS to a stiffer one which corresponds to a shift from DM core to DM halo configuration. Note that the DM core - halo transition is specified by the ratio of $R_{B}/R_{D}$ while the increasing of $M_{T_{max}}$ and $\Lambda_{1.4}$, which is due to the  DM halo formation, occurs at relatively higher DM fractions where the EoS is stiff enough. 

Furthermore, as the main objective of this research, by comparing both of the Figures \ref{MT-FX} and \ref{Lam-FX}, it is shown that for  $4\%\lesssim F_{\chi}\lesssim20\%$, massive bosons, $m_{\chi}\gtrsim200$, and small coupling constants, $\lambda\lesssim2\pi$, present a bosonic DM parameter space for which both $M_{T_{max}}$ and $\Lambda_{1.4}$ are consistent with the astrophysical bounds. This result indicates the modification of NS quantities resulting from DD2 EoS by involving self-interacting bosonic DM.

\section{Conclusion}
\label{sec6}

The occurrence of self-repulsive bosonic DM has been taken into account to investigate the impact of DM on NS features for which the BM is described by the well-known DD2 model. We have seen that for low DM fractions, a dense core is mainly formed inside NSs while by increasing the fraction, bosonic particles distribute as a halo around the NS. It is shown that for massive bosons and low coupling constants,  DM will reside as a core for wider ranges of $F_{\chi}$, however, for light bosons and large values of $\lambda$, an extended DM halo will be formed for the majority of the considered DM fractions. Moreover, for each applied DM model parameter a transition occurs from DM core to DM halo formation by varying the
parameter $F_{\chi}$.

Furthermore, we show that a DM core reduces the maximum mass, radius and tidal deformability, but a DM halo will enhance all of them. Therefore, these features demonstrate that including DM in NSs can modify the observable quantities induced from the considered nuclear matter model for the NS interior. Owing to the point that DD2 EoS provides a value for $\Lambda_{1.4}$ which is not consistent with the current observational bound, we examine the bosonic DM parameter space so as to find the regions that make the observable features ($M_{T_{max}}$ and $\Lambda_{1.4}$) compatible with their astrophysical limits. We have indicated that for massive bosons ($m_{\chi}\gtrsim200$) and small coupling constants ($\lambda\lesssim2\pi$), DM fractions within the range of 4\% and  20\%,  are more favorable and improve the tidal deformability so that the results are consistent with both $M_{max}$ and $\Lambda_{1.4}$ constraints. There are no uncertainties given for the DD2 model and its parameterization which was adjusted at and below saturation density based on the properties of atomic nuclei. A modification of this parameterization to draw new conclusions is out of the scope of the present work and will be deferred to a further study since it requires a careful consideration of parameter correlations. 

Finally, by considering the great possibility of accumulation of DM in NSs which alters its observable quantities, future work will account for a larger number of nuclear matter EoSs. Indeed, an overlap of allowed DM parameter spaces can be obtained for the collection of BM EoSs in which all of the observable features are compatible with their constraints. It is noteworthy to say that the analysis performed in this research shows the potential of DM in relaxing the uncertainties in the baryonic EoS space. Furthermore, regarding the advanced GW detectors and X-ray and radio telescopes, conducting similar investigations will help to shed more light on high-density matter in NSs and even the nature of DM.

\vspace{6pt}

\authorcontributions{D.R.K. developing the idea and DM model, investigation, methodology, analysis and calculations and preparing the original draft, M. Sh. producing baryonic matter EoS data, consultation for presenting plots and drafting the paper, reviewing and editing, S.S. developing the idea and DM model, consultation for drafting the paper, reviewing and editing,  S.T. obtaining and programming the
baryonic matter EoS, reviewing and editing. All authors have contributed to discussions, read and agreed to the published 
version of the manuscript.}

\funding{This research received no external funding}

\institutionalreview{Not applicable}

\dataavailability{Not applicable}

\acknowledgments{D.R.K expresses gratitude to Violetta Sagun and Oleksii Ivanytskyi for their support in presenting this work at the "Dark Matter and Stars" conference and for engaging in other discussions.
M. Sh. has been supported by the program Excellence Initiative--Research University of the University of Wroclaw of the Ministry of Education and Science. M. Sh. acknowledges the hospitality of the Frankfurt Institute for Advanced Studies (FIAS).}

\conflictsofinterest{The authors declare no conflict of interest.}

\abbreviations{Abbreviations}{
The following abbreviations are used in this manuscript:\\

\noindent 

\begin{description}
\label{tab:abbreviations}
\item[NS]{Neutron Star}
\item[DM]{Dark Matter} 
\item[BM]{Baryonic Matter}
\item[EoS]{Equation of State}
\item[TOV]{Tolman-Oppenheimer-Volkof}
\item[DD2]{Parameterization of a GDRF for hadronic matter including only nucleons}
\item[GW]{Gravitational-Wave}
\item[NICER]{Neutron Star Interior Composition ExploreR}

\end{description}
}

\begin{adjustwidth}{-\extralength}{0cm}

\reftitle{References}

\bibliography{particles-2776285}

\begin{thebibliography}{999}

\bibitem[Baryakhtar et~al.(2022)]{Baryakhtar:2022hbu}
Baryakhtar, M.;  et~al.
\newblock {Dark Matter In Extreme Astrophysical Environments}.
\newblock In Proceedings of the {2022 Snowmass Summer Study},  3 2022,
  \href{http://xxx.lanl.gov/abs/2203.07984}{{\normalfont
  [arXiv:hep-ph/2203.07984]}}.

\bibitem[Leane and Smirnov(2021)]{Leane:2020wob}
Leane, R.K.; Smirnov, J.
\newblock {Exoplanets as Sub-GeV Dark Matter Detectors}.
\newblock {\em Phys. Rev. Lett.} {\bf 2021}, {\em 126},~161101,
  \href{http://xxx.lanl.gov/abs/2010.00015}{{\normalfont
  [arXiv:hep-ph/2010.00015]}}.
\newblock {\url{https://doi.org/10.1103/PhysRevLett.126.161101}}.

\bibitem[Bramante and Raj(2023)]{Bramante:2023djs}
Bramante, J.; Raj, N.
\newblock {Dark matter in compact stars} {\bf 2023}.
\newblock  \href{http://xxx.lanl.gov/abs/2307.14435}{{\normalfont
  [arXiv:hep-ph/2307.14435]}}.

\bibitem[Leane and Smirnov(2023)]{Leane:2022hkk}
Leane, R.K.; Smirnov, J.
\newblock {Floating dark matter in celestial bodies}.
\newblock {\em JCAP} {\bf 2023}, {\em 10},~057,
  \href{http://xxx.lanl.gov/abs/2209.09834}{{\normalfont
  [arXiv:hep-ph/2209.09834]}}.
\newblock {\url{https://doi.org/10.1088/1475-7516/2023/10/057}}.

\bibitem[Ellis et~al.(2018)Ellis, Hütsi, Kannike, Marzola, Raidal, and
  Vaskonen]{Ellis:2018bkr}
Ellis, J.; Hütsi, G.; Kannike, K.; Marzola, L.; Raidal, M.; Vaskonen, V.
\newblock {Dark Matter Effects On Neutron Star Properties}.
\newblock {\em Phys. Rev. D} {\bf 2018}, {\em 97},~123007,
  \href{http://xxx.lanl.gov/abs/1804.01418}{{\normalfont
  [arXiv:astro-ph.CO/1804.01418]}}.
\newblock {\url{https://doi.org/10.1103/PhysRevD.97.123007}}.

\bibitem[Nelson et~al.(2019)Nelson, Reddy, and Zhou]{Nelson:2018xtr}
Nelson, A.; Reddy, S.; Zhou, D.
\newblock {Dark halos around neutron stars and gravitational waves}.
\newblock {\em JCAP} {\bf 2019}, {\em 07},~012,
  \href{http://xxx.lanl.gov/abs/1803.03266}{{\normalfont
  [arXiv:hep-ph/1803.03266]}}.
\newblock {\url{https://doi.org/10.1088/1475-7516/2019/07/012}}.

\bibitem[Ryan and Radice(2022)]{Ryan:2022hku}
Ryan, M.; Radice, D.
\newblock {Exotic compact objects: The dark white dwarf}.
\newblock {\em Phys. Rev. D} {\bf 2022}, {\em 105},~115034,
  \href{http://xxx.lanl.gov/abs/2201.05626}{{\normalfont
  [arXiv:astro-ph.HE/2201.05626]}}.
\newblock {\url{https://doi.org/10.1103/PhysRevD.105.115034}}.

\bibitem[Chan et~al.(2022)Chan, Chu, and Leung]{Chan:2021gcm}
Chan, H.S.; Chu, M.c.; Leung, S.C.
\newblock {Dark Matter\textendash{}admixed Rotating White Dwarfs as Peculiar
  Compact Objects}.
\newblock {\em Astrophys. J.} {\bf 2022}, {\em 941},~115,
  \href{http://xxx.lanl.gov/abs/2111.12894}{{\normalfont
  [arXiv:astro-ph.HE/2111.12894]}}.
\newblock {\url{https://doi.org/10.3847/1538-4357/aca09b}}.

\bibitem[Liang and Shao(2023)]{Liang:2023nvo}
Liang, D.; Shao, L.
\newblock {Improved bounds on the bosonic dark matter with pulsars in the Milky
  Way}.
\newblock {\em JCAP} {\bf 2023}, {\em 08},~016,
  \href{http://xxx.lanl.gov/abs/2303.05107}{{\normalfont
  [arXiv:astro-ph.HE/2303.05107]}}.
\newblock {\url{https://doi.org/10.1088/1475-7516/2023/08/016}}.

\bibitem[Karkevandi et~al.(2022)Karkevandi, Shakeri, Sagun, and
  Ivanytskyi]{Karkevandi:2021ygv}
Karkevandi, D.R.; Shakeri, S.; Sagun, V.; Ivanytskyi, O.
\newblock {Bosonic dark matter in neutron stars and its effect on gravitational
  wave signal}.
\newblock {\em Phys. Rev. D} {\bf 2022}, {\em 105},~023001,
  \href{http://xxx.lanl.gov/abs/2109.03801}{{\normalfont
  [arXiv:astro-ph.HE/2109.03801]}}.
\newblock {\url{https://doi.org/10.1103/PhysRevD.105.023001}}.

\bibitem[Shakeri and Karkevandi(2022)]{Shakeri:2022dwg}
Shakeri, S.; Karkevandi, D.R.
\newblock {Bosonic Dark Matter in Light of the NICER Precise Mass-Radius
  Measurements} {\bf 2022}.
\newblock  \href{http://xxx.lanl.gov/abs/2210.17308}{{\normalfont
  [arXiv:astro-ph.HE/2210.17308]}}.

\bibitem[Diedrichs et~al.(2023)Diedrichs, Becker, Jockel, Christian, Sagunski,
  and Schaffner-Bielich]{Diedrichs:2023trk}
Diedrichs, R.F.; Becker, N.; Jockel, C.; Christian, J.E.; Sagunski, L.;
  Schaffner-Bielich, J.
\newblock {Tidal deformability of fermion-boson stars: Neutron stars admixed
  with ultralight dark matter}.
\newblock {\em Phys. Rev. D} {\bf 2023}, {\em 108},~064009,
  \href{http://xxx.lanl.gov/abs/2303.04089}{{\normalfont
  [arXiv:gr-qc/2303.04089]}}.
\newblock {\url{https://doi.org/10.1103/PhysRevD.108.064009}}.

\bibitem[Rutherford et~al.(2023)Rutherford, Raaijmakers, Prescod-Weinstein, and
  Watts]{Rutherford:2022xeb}
Rutherford, N.; Raaijmakers, G.; Prescod-Weinstein, C.; Watts, A.
\newblock {Constraining bosonic asymmetric dark matter with neutron star
  mass-radius measurements}.
\newblock {\em Phys. Rev. D} {\bf 2023}, {\em 107},~103051,
  \href{http://xxx.lanl.gov/abs/2208.03282}{{\normalfont
  [arXiv:astro-ph.HE/2208.03282]}}.
\newblock {\url{https://doi.org/10.1103/PhysRevD.107.103051}}.

\bibitem[Giangrandi et~al.(2023)Giangrandi, Sagun, Ivanytskyi, Provid\^encia,
  and Dietrich]{Giangrandi:2022wht}
Giangrandi, E.; Sagun, V.; Ivanytskyi, O.; Provid\^encia, C.; Dietrich, T.
\newblock {The Effects of Self-interacting Bosonic Dark Matter on Neutron Star
  Properties}.
\newblock {\em Astrophys. J.} {\bf 2023}, {\em 953},~115,
  \href{http://xxx.lanl.gov/abs/2209.10905}{{\normalfont
  [arXiv:astro-ph.HE/2209.10905]}}.
\newblock {\url{https://doi.org/10.3847/1538-4357/ace104}}.

\bibitem[Ivanytskyi et~al.(2020)Ivanytskyi, Sagun, and
  Lopes]{PhysRevD.102.063028}
Ivanytskyi, O.; Sagun, V.; Lopes, I.
\newblock Neutron stars: New constraints on asymmetric dark matter.
\newblock {\em Phys. Rev. D} {\bf 2020}, {\em 102},~063028.
\newblock {\url{https://doi.org/10.1103/PhysRevD.102.063028}}.

\bibitem[Deliyergiyev et~al.(2023)Deliyergiyev, Del~Popolo, and
  Delliou]{Deliyergiyev:2023uer}
Deliyergiyev, M.; Del~Popolo, A.; Delliou, M.L.
\newblock {Neutron star mass in dark matter clumps} {\bf 2023}.
\newblock  \href{http://xxx.lanl.gov/abs/2311.00113}{{\normalfont
  [arXiv:astro-ph.GA/2311.00113]}}.
\newblock {\url{https://doi.org/10.1093/mnras/stad3311}}.

\bibitem[Bhattacharya et~al.(2023)Bhattacharya, Dasgupta, Laha, and
  Ray]{Bhattacharya:2023stq}
Bhattacharya, S.; Dasgupta, B.; Laha, R.; Ray, A.
\newblock {Can LIGO Detect Nonannihilating Dark Matter?}
\newblock {\em Phys. Rev. Lett.} {\bf 2023}, {\em 131},~091401,
  \href{http://xxx.lanl.gov/abs/2302.07898}{{\normalfont
  [arXiv:hep-ph/2302.07898]}}.
\newblock {\url{https://doi.org/10.1103/PhysRevLett.131.091401}}.

\bibitem[Shahrbaf(2023)]{Shahrbaf:2023uxy}
Shahrbaf, M.
\newblock {Appearance of sexaquark in the core of neutron stars as a candidate
  of dark matter}.
\newblock {\em J. Phys. Conf. Ser.} {\bf 2023}, {\em 2536},~012001.
\newblock {\url{https://doi.org/10.1088/1742-6596/2536/1/012001}}.

\bibitem[Shahrbaf et~al.(2022)Shahrbaf, Blaschke, Typel, Farrar, and
  Alvarez-Castillo]{Shahrbaf:2022upc}
Shahrbaf, M.; Blaschke, D.; Typel, S.; Farrar, G.R.; Alvarez-Castillo, D.E.
\newblock {Sexaquark dilemma in neutron stars and its solution by quark
  deconfinement}.
\newblock {\em Phys. Rev. D} {\bf 2022}, {\em 105},~103005,
  \href{http://xxx.lanl.gov/abs/2202.00652}{{\normalfont
  [arXiv:nucl-th/2202.00652]}}.
\newblock {\url{https://doi.org/10.1103/PhysRevD.105.103005}}.

\bibitem[Berryman et~al.(2022)Berryman, Gardner, and Zakeri]{Berryman:2022zic}
Berryman, J.M.; Gardner, S.; Zakeri, M.
\newblock {Neutron Stars with Baryon Number Violation, Probing Dark Sectors}.
\newblock {\em Symmetry} {\bf 2022}, {\em 14},~518,
  \href{http://xxx.lanl.gov/abs/2201.02637}{{\normalfont
  [arXiv:hep-ph/2201.02637]}}.
\newblock {\url{https://doi.org/10.3390/sym14030518}}.

\bibitem[Shirke et~al.(2023)Shirke, Ghosh, Chatterjee, Sagunski, and
  Schaffner-Bielich]{Shirke:2023ktu}
Shirke, S.; Ghosh, S.; Chatterjee, D.; Sagunski, L.; Schaffner-Bielich, J.
\newblock {R-modes as a New Probe of Dark Matter in Neutron Stars} {\bf 2023}.
\newblock  \href{http://xxx.lanl.gov/abs/2305.05664}{{\normalfont
  [arXiv:astro-ph.HE/2305.05664]}}.

\bibitem[Husain et~al.(2022)Husain, Motta, and Thomas]{Husain:2022bxl}
Husain, W.; Motta, T.F.; Thomas, A.W.
\newblock {Consequences of neutron decay inside neutron stars}.
\newblock {\em JCAP} {\bf 2022}, {\em 10},~028,
  \href{http://xxx.lanl.gov/abs/2203.02758}{{\normalfont
  [arXiv:hep-ph/2203.02758]}}.
\newblock {\url{https://doi.org/10.1088/1475-7516/2022/10/028}}.

\bibitem[Maselli et~al.(2017)Maselli, Pnigouras, Nielsen, Kouvaris, and
  Kokkotas]{Maselli_2017}
Maselli, A.; Pnigouras, P.; Nielsen, N.G.; Kouvaris, C.; Kokkotas, K.D.
\newblock Dark stars: Gravitational and electromagnetic observables.
\newblock {\em Physical Review D} {\bf 2017}, {\em 96}.
\newblock {\url{https://doi.org/10.1103/physrevd.96.023005}}.

\bibitem[Pitz and Schaffner-Bielich(2023)]{Pitz:2023ejc}
Pitz, S.L.; Schaffner-Bielich, J.
\newblock {Generating ultra compact boson stars with modified scalar
  potentials} {\bf 2023}.
\newblock  \href{http://xxx.lanl.gov/abs/2308.01254}{{\normalfont
  [arXiv:astro-ph.HE/2308.01254]}}.

\bibitem[Cassing et~al.(2023)Cassing, Brisebois, Azeem, and
  Schaffner-Bielich]{Cassing:2022tnn}
Cassing, M.; Brisebois, A.; Azeem, M.; Schaffner-Bielich, J.
\newblock {Exotic Compact Objects with Two Dark Matter Fluids}.
\newblock {\em Astrophys. J.} {\bf 2023}, {\em 944},~130,
  \href{http://xxx.lanl.gov/abs/2210.13697}{{\normalfont
  [arXiv:gr-qc/2210.13697]}}.
\newblock {\url{https://doi.org/10.3847/1538-4357/acb3be}}.

\bibitem[Eby et~al.(2016)Eby, Kouvaris, Nielsen, and Wijewardhana]{Eby:2015hsq}
Eby, J.; Kouvaris, C.; Nielsen, N.G.; Wijewardhana, L.
\newblock {Boson Stars from Self-Interacting Dark Matter}.
\newblock {\em JHEP} {\bf 2016}, {\em 02},~028,
  \href{http://xxx.lanl.gov/abs/1511.04474}{{\normalfont
  [arXiv:hep-ph/1511.04474]}}.
\newblock {\url{https://doi.org/10.1007/JHEP02(2016)028}}.

\bibitem[Dietrich et~al.(2019)Dietrich, Day, Clough, Coughlin, and
  Niemeyer]{Dietrich:2018jov}
Dietrich, T.; Day, F.; Clough, K.; Coughlin, M.; Niemeyer, J.
\newblock {Neutron star\textendash{}axion star collisions in the light of
  multimessenger astronomy}.
\newblock {\em Mon. Not. Roy. Astron. Soc.} {\bf 2019}, {\em 483},~908--914,
  \href{http://xxx.lanl.gov/abs/1808.04746}{{\normalfont
  [arXiv:astro-ph.HE/1808.04746]}}.
\newblock {\url{https://doi.org/10.1093/mnras/sty3158}}.

\bibitem[Clough et~al.(2018)Clough, Dietrich, and Niemeyer]{Clough:2018exo}
Clough, K.; Dietrich, T.; Niemeyer, J.C.
\newblock {Axion star collisions with black holes and neutron stars in full 3D
  numerical relativity}.
\newblock {\em Phys. Rev. D} {\bf 2018}, {\em 98},~083020,
  \href{http://xxx.lanl.gov/abs/1808.04668}{{\normalfont
  [arXiv:gr-qc/1808.04668]}}.
\newblock {\url{https://doi.org/10.1103/PhysRevD.98.083020}}.

\bibitem[Huth et~al.(2022)]{Huth:2021bsp}
Huth, S.;  et~al.
\newblock {Constraining Neutron-Star Matter with Microscopic and Macroscopic
  Collisions}.
\newblock {\em Nature} {\bf 2022}, {\em 606},~276--280,
  \href{http://xxx.lanl.gov/abs/2107.06229}{{\normalfont
  [arXiv:nucl-th/2107.06229]}}.
\newblock {\url{https://doi.org/10.1038/s41586-022-04750-w}}.

\bibitem[Raaijmakers et~al.(2021)Raaijmakers, Greif, Hebeler, Hinderer,
  Nissanke, Schwenk, Riley, Watts, Lattimer, and Ho]{Raaijmakers:2021uju}
Raaijmakers, G.; Greif, S.K.; Hebeler, K.; Hinderer, T.; Nissanke, S.; Schwenk,
  A.; Riley, T.E.; Watts, A.L.; Lattimer, J.M.; Ho, W.C.G.
\newblock {Constraints on the Dense Matter Equation of State and Neutron Star
  Properties from NICER\textquoteright{}s Mass\textendash{}Radius Estimate of
  PSR J0740+6620 and Multimessenger Observations}.
\newblock {\em Astrophys. J. Lett.} {\bf 2021}, {\em 918},~L29,
  \href{http://xxx.lanl.gov/abs/2105.06981}{{\normalfont
  [arXiv:astro-ph.HE/2105.06981]}}.
\newblock {\url{https://doi.org/10.3847/2041-8213/ac089a}}.

\bibitem[Hippert et~al.(2023)Hippert, Dillingham, Tan, Curtin, Noronha-Hostler,
  and Yunes]{Hippert:2022snq}
Hippert, M.; Dillingham, E.; Tan, H.; Curtin, D.; Noronha-Hostler, J.; Yunes,
  N.
\newblock {Dark matter or regular matter in neutron stars? How to tell the
  difference from the coalescence of compact objects}.
\newblock {\em Phys. Rev. D} {\bf 2023}, {\em 107},~115028,
  \href{http://xxx.lanl.gov/abs/2211.08590}{{\normalfont
  [arXiv:astro-ph.HE/2211.08590]}}.
\newblock {\url{https://doi.org/10.1103/PhysRevD.107.115028}}.

\bibitem[Collier et~al.(2022)Collier, Croon, and Leane]{Collier:2022cpr}
Collier, M.; Croon, D.; Leane, R.K.
\newblock {Tidal Love numbers of novel and admixed celestial objects}.
\newblock {\em Phys. Rev. D} {\bf 2022}, {\em 106},~123027,
  \href{http://xxx.lanl.gov/abs/2205.15337}{{\normalfont
  [arXiv:gr-qc/2205.15337]}}.
\newblock {\url{https://doi.org/10.1103/PhysRevD.106.123027}}.

\bibitem[Rafiei~Karkevandi et~al.(2021)Rafiei~Karkevandi, Shakeri, Sagun, and
  Ivanytskyi]{RafieiKarkevandi:2021hcc}
Rafiei~Karkevandi, D.; Shakeri, S.; Sagun, V.; Ivanytskyi, O.
\newblock {Tidal deformability as a probe of dark matter in neutron stars}.
\newblock In Proceedings of the {16th Marcel Grossmann Meeting on~Recent
  Developments in Theoretical and Experimental General Relativity, Astrophysics
  and Relativistic Field Theories},  12 2021,
  \href{http://xxx.lanl.gov/abs/2112.14231}{{\normalfont
  [arXiv:astro-ph.HE/2112.14231]}}.
\newblock {\url{https://doi.org/10.1142/9789811269776_0307}}.

\bibitem[Dengler et~al.(2022)Dengler, Schaffner-Bielich, and
  Tolos]{Dengler:2021qcq}
Dengler, Y.; Schaffner-Bielich, J.; Tolos, L.
\newblock {Second Love number of dark compact planets and neutron stars with
  dark matter}.
\newblock {\em Phys. Rev. D} {\bf 2022}, {\em 105},~043013,
  \href{http://xxx.lanl.gov/abs/2111.06197}{{\normalfont
  [arXiv:astro-ph.HE/2111.06197]}}.
\newblock {\url{https://doi.org/10.1103/PhysRevD.105.043013}}.

\bibitem[Routaray et~al.(2023)Routaray, Das, Sen, Kumar, Panotopoulos, and
  Zhao]{Routaray:2022utr}
Routaray, P.; Das, H.C.; Sen, S.; Kumar, B.; Panotopoulos, G.; Zhao, T.
\newblock {Radial oscillations of dark matter admixed neutron stars}.
\newblock {\em Phys. Rev. D} {\bf 2023}, {\em 107},~103039,
  \href{http://xxx.lanl.gov/abs/2211.12808}{{\normalfont
  [arXiv:nucl-th/2211.12808]}}.
\newblock {\url{https://doi.org/10.1103/PhysRevD.107.103039}}.

\bibitem[Cronin et~al.(2023)Cronin, Zhang, and Kain]{Cronin:2023xzc}
Cronin, J.; Zhang, X.; Kain, B.
\newblock {Rotating dark matter admixed neutron stars}.
\newblock {\em Phys. Rev. D} {\bf 2023}, {\em 108},~103016,
  \href{http://xxx.lanl.gov/abs/2311.07714}{{\normalfont
  [arXiv:gr-qc/2311.07714]}}.
\newblock {\url{https://doi.org/10.1103/PhysRevD.108.103016}}.

\bibitem[Jockel and Sagunski(2023)]{Jockel:2023rrm}
Jockel, C.; Sagunski, L.
\newblock {Fermion Proca Stars: Vector Dark Matter Admixed Neutron Stars} {\bf
  2023}.
\newblock  \href{http://xxx.lanl.gov/abs/2310.17291}{{\normalfont
  [arXiv:gr-qc/2310.17291]}}.

\bibitem[Panotopoulos and Lopes(2017)]{Panotopoulos_2017}
Panotopoulos, G.; Lopes, I.
\newblock Dark matter effect on realistic equation of state in neutron stars.
\newblock {\em Physical Review D} {\bf 2017}, {\em 96}.
\newblock {\url{https://doi.org/10.1103/physrevd.96.083004}}.

\bibitem[Das et~al.(2020)Das, Kumar, Kumar, Kumar~Biswal, Nakatsukasa, Li, and
  Patra]{Das:2020vng}
Das, H.C.; Kumar, A.; Kumar, B.; Kumar~Biswal, S.; Nakatsukasa, T.; Li, A.;
  Patra, S.K.
\newblock {Effects of dark matter on the nuclear and neutron star matter}.
\newblock {\em Mon. Not. Roy. Astron. Soc.} {\bf 2020}, {\em 495},~4893--4903,
  \href{http://xxx.lanl.gov/abs/2002.00594}{{\normalfont
  [arXiv:nucl-th/2002.00594]}}.
\newblock {\url{https://doi.org/10.1093/mnras/staa1435}}.

\bibitem[Louren\c{c}o et~al.(2022)Louren\c{c}o, Lenzi, Frederico, and
  Dutra]{Lourenco:2022fmf}
Louren\c{c}o, O.; Lenzi, C.H.; Frederico, T.; Dutra, M.
\newblock {Dark matter effects on tidal deformabilities and moment of inertia
  in a hadronic model with short-range correlations}.
\newblock {\em Phys. Rev. D} {\bf 2022}, {\em 106},~043010,
  \href{http://xxx.lanl.gov/abs/2208.06067}{{\normalfont
  [arXiv:nucl-th/2208.06067]}}.
\newblock {\url{https://doi.org/10.1103/PhysRevD.106.043010}}.

\bibitem[Routaray et~al.(2023)Routaray, Mohanty, Das, Ghosh, Kalita, Parmar,
  and Kumar]{Routaray:2023spb}
Routaray, P.; Mohanty, S.R.; Das, H.C.; Ghosh, S.; Kalita, P.J.; Parmar, V.;
  Kumar, B.
\newblock {Investigating Dark Matter-Admixed Neutron Stars with NITR Equation
  of State in Light of PSR J0952-0607}.
\newblock {\em JCAP} {\bf 2023}, {\em 10},~073,
  \href{http://xxx.lanl.gov/abs/2304.05100}{{\normalfont
  [arXiv:nucl-th/2304.05100]}}.
\newblock {\url{https://doi.org/10.1088/1475-7516/2023/10/073}}.

\bibitem[Guha and Sen(2024)]{Guha:2024pnn}
Guha, A.; Sen, D.
\newblock {Constraining the mass of fermionic dark matter from its feeble
  interaction with hadronic matter via dark mediators in neutron stars} {\bf
  2024}.
\newblock  \href{http://xxx.lanl.gov/abs/2401.14419}{{\normalfont
  [arXiv:astro-ph.HE/2401.14419]}}.

\bibitem[Sandin and Ciarcelluti(2009)]{Sandin_2009}
Sandin, F.; Ciarcelluti, P.
\newblock Effects of mirror dark matter on neutron stars.
\newblock {\em Astroparticle Physics} {\bf 2009}, {\em 32},~278–284.
\newblock {\url{https://doi.org/10.1016/j.astropartphys.2009.09.005}}.

\bibitem[Ciarcelluti and Sandin(2011)]{Ciarcelluti:2010ji}
Ciarcelluti, P.; Sandin, F.
\newblock {Have neutron stars a dark matter core?}
\newblock {\em Phys. Lett. B} {\bf 2011}, {\em 695},~19--21,
  \href{http://xxx.lanl.gov/abs/1005.0857}{{\normalfont
  [arXiv:astro-ph.HE/1005.0857]}}.
\newblock {\url{https://doi.org/10.1016/j.physletb.2010.11.021}}.

\bibitem[Rezaei(2023)]{Rezaei:2023iif}
Rezaei, Z.
\newblock {Fuzzy dark matter in relativistic stars}.
\newblock {\em Mon. Not. Roy. Astron. Soc.} {\bf 2023}, {\em 524},~2015--2024,
  \href{http://xxx.lanl.gov/abs/2306.17665}{{\normalfont
  [arXiv:astro-ph.HE/2306.17665]}}.
\newblock {\url{https://doi.org/10.1093/mnras/stad1975}}.

\bibitem[Thakur et~al.(2023)Thakur, Malik, Das, Jha, and
  Provid\^encia]{Thakur:2023aqm}
Thakur, P.; Malik, T.; Das, A.; Jha, T.K.; Provid\^encia, C.
\newblock {Exploring robust correlations between fermionic dark matter model
  parameters and neutron star properties: A two-fluid perspective} {\bf 2023}.
\newblock  \href{http://xxx.lanl.gov/abs/2308.00650}{{\normalfont
  [arXiv:hep-ph/2308.00650]}}.

\bibitem[Rezaei(2017)]{Rezaei_2017}
Rezaei, Z.
\newblock STUDY OF DARK-MATTER ADMIXED NEUTRON STARS USING THE EQUATION OF
  STATE FROM THE ROTATIONAL CURVES OF GALAXIES.
\newblock {\em The Astrophysical Journal} {\bf 2017}, {\em 835},~33.
\newblock {\url{https://doi.org/10.1088/1361-6528/aa5273}}.

\bibitem[Gleason et~al.(2022)Gleason, Brown, and Kain]{Gleason:2022eeg}
Gleason, T.; Brown, B.; Kain, B.
\newblock {Dynamical evolution of dark matter admixed neutron stars}.
\newblock {\em Phys. Rev. D} {\bf 2022}, {\em 105},~023010,
  \href{http://xxx.lanl.gov/abs/2201.02274}{{\normalfont
  [arXiv:gr-qc/2201.02274]}}.
\newblock {\url{https://doi.org/10.1103/PhysRevD.105.023010}}.

\bibitem[Sun and Wen(2023)]{Sun:2023cqr}
Sun, H.; Wen, D.
\newblock {A new criterion for the existence of dark matter in neutron stars}
  {\bf 2023}.
\newblock  \href{http://xxx.lanl.gov/abs/2312.17288}{{\normalfont
  [arXiv:astro-ph.HE/2312.17288]}}.

\bibitem[Kaplan et~al.(2009)Kaplan, Luty, and Zurek]{Kaplan:2009ag}
Kaplan, D.E.; Luty, M.A.; Zurek, K.M.
\newblock {Asymmetric Dark Matter}.
\newblock {\em Phys. Rev. D} {\bf 2009}, {\em 79},~115016,
  \href{http://xxx.lanl.gov/abs/0901.4117}{{\normalfont
  [arXiv:hep-ph/0901.4117]}}.
\newblock {\url{https://doi.org/10.1103/PhysRevD.79.115016}}.

\bibitem[\'Avila et~al.(2023)\'Avila, Giangrandi, Sagun, Ivanytskyi, and
  Provid\^encia]{Avila:2023rzj}
\'Avila, A.; Giangrandi, E.; Sagun, V.; Ivanytskyi, O.; Provid\^encia, C.
\newblock {Rapid neutron star cooling triggered by accumulated dark matter}
  {\bf 2023}.
\newblock  \href{http://xxx.lanl.gov/abs/2309.03894}{{\normalfont
  [arXiv:astro-ph.HE/2309.03894]}}.

\bibitem[\'Angeles P\'erez-Garc\'\i{}a et~al.(2022)\'Angeles
  P\'erez-Garc\'\i{}a, Grigorian, Albertus, Barba, and
  Silk]{AngelesPerez-Garcia:2022qzs}
\'Angeles P\'erez-Garc\'\i{}a, M.; Grigorian, H.; Albertus, C.; Barba, D.;
  Silk, J.
\newblock {Cooling of Neutron Stars admixed with light dark matter: A case
  study}.
\newblock {\em Phys. Lett. B} {\bf 2022}, {\em 827},~136937,
  \href{http://xxx.lanl.gov/abs/2202.00702}{{\normalfont
  [arXiv:hep-ph/2202.00702]}}.
\newblock {\url{https://doi.org/10.1016/j.physletb.2022.136937}}.

\bibitem[Chatterjee et~al.(2022)Chatterjee, Garani, Jain, Kanodia, Kumar, and
  Vempati]{Chatterjee:2022dhp}
Chatterjee, S.; Garani, R.; Jain, R.K.; Kanodia, B.; Kumar, M.S.N.; Vempati,
  S.K.
\newblock {Faint light of old neutron stars from dark matter capture and
  detectability at the James Webb Space Telescope} {\bf 2022}.
\newblock  \href{http://xxx.lanl.gov/abs/2205.05048}{{\normalfont
  [arXiv:astro-ph.HE/2205.05048]}}.

\bibitem[Alvarez et~al.(2023)Alvarez, Joglekar, Phoroutan-Mehr, and
  Yu]{Alvarez:2023fjj}
Alvarez, G.; Joglekar, A.; Phoroutan-Mehr, M.; Yu, H.B.
\newblock {Heating neutron stars with inelastic dark matter and relativistic
  targets}.
\newblock {\em Phys. Rev. D} {\bf 2023}, {\em 107},~103024,
  \href{http://xxx.lanl.gov/abs/2301.08767}{{\normalfont
  [arXiv:hep-ph/2301.08767]}}.
\newblock {\url{https://doi.org/10.1103/PhysRevD.107.103024}}.

\bibitem[Nguyen and Tait(2023)]{Nguyen:2022zwb}
Nguyen, T.T.Q.; Tait, T.M.P.
\newblock {Bounds on long-lived dark matter mediators from neutron stars}.
\newblock {\em Phys. Rev. D} {\bf 2023}, {\em 107},~115016,
  \href{http://xxx.lanl.gov/abs/2212.12547}{{\normalfont
  [arXiv:hep-ph/2212.12547]}}.
\newblock {\url{https://doi.org/10.1103/PhysRevD.107.115016}}.

\bibitem[Bauswein et~al.(2023)Bauswein, Guo, Lien, Lin, and
  Wu]{Bauswein:2020kor}
Bauswein, A.; Guo, G.; Lien, J.H.; Lin, Y.H.; Wu, M.R.
\newblock {Compact dark objects in neutron star mergers}.
\newblock {\em Phys. Rev. D} {\bf 2023}, {\em 107},~083002,
  \href{http://xxx.lanl.gov/abs/2012.11908}{{\normalfont
  [arXiv:astro-ph.HE/2012.11908]}}.
\newblock {\url{https://doi.org/10.1103/PhysRevD.107.083002}}.

\bibitem[Bezares et~al.(2019)Bezares, Vigan\`o, and
  Palenzuela]{Bezares:2019jcb}
Bezares, M.; Vigan\`o, D.; Palenzuela, C.
\newblock {Gravitational wave signatures of dark matter cores in binary neutron
  star mergers by using numerical simulations}.
\newblock {\em Phys. Rev. D} {\bf 2019}, {\em 100},~044049,
  \href{http://xxx.lanl.gov/abs/1905.08551}{{\normalfont
  [arXiv:gr-qc/1905.08551]}}.
\newblock {\url{https://doi.org/10.1103/PhysRevD.100.044049}}.

\bibitem[Ellis et~al.(2018)Ellis, Hektor, Hütsi, Kannike, Marzola, Raidal, and
  Vaskonen]{Ellis:2017jgp}
Ellis, J.; Hektor, A.; Hütsi, G.; Kannike, K.; Marzola, L.; Raidal, M.;
  Vaskonen, V.
\newblock {Search for Dark Matter Effects on Gravitational Signals from Neutron
  Star Mergers}.
\newblock {\em Phys. Lett. B} {\bf 2018}, {\em 781},~607--610,
  \href{http://xxx.lanl.gov/abs/1710.05540}{{\normalfont
  [arXiv:astro-ph.CO/1710.05540]}}.
\newblock {\url{https://doi.org/10.1016/j.physletb.2018.04.048}}.

\bibitem[Emma et~al.(2022)Emma, Schianchi, Pannarale, Sagun, and
  Dietrich]{Emma:2022xjs}
Emma, M.; Schianchi, F.; Pannarale, F.; Sagun, V.; Dietrich, T.
\newblock {Numerical Simulations of Dark Matter Admixed Neutron Star Binaries}.
\newblock {\em Particles} {\bf 2022}, {\em 5},~273--286,
  \href{http://xxx.lanl.gov/abs/2206.10887}{{\normalfont
  [arXiv:gr-qc/2206.10887]}}.
\newblock {\url{https://doi.org/10.3390/particles5030024}}.

\bibitem[R\"uter et~al.(2023)R\"uter, Sagun, Tichy, and
  Dietrich]{Ruter:2023uzc}
R\"uter, H.R.; Sagun, V.; Tichy, W.; Dietrich, T.
\newblock {Quasi-equilibrium configurations of binary systems of dark matter
  admixed neutron stars} {\bf 2023}.
\newblock  \href{http://xxx.lanl.gov/abs/2301.03568}{{\normalfont
  [arXiv:gr-qc/2301.03568]}}.

\bibitem[Das et~al.(2021)Das, Kumar, and Patra]{Das:2021wku}
Das, H.C.; Kumar, A.; Patra, S.K.
\newblock {Effects of dark matter on the inspiral properties of the binary
  neutron star} {\bf 2021}.
\newblock  \href{http://xxx.lanl.gov/abs/2104.01815}{{\normalfont
  [arXiv:astro-ph.HE/2104.01815]}}.

\bibitem[Miller et~al.(2019)]{Miller:2019cac}
Miller, M.C.;  et~al.
\newblock {PSR J0030+0451 Mass and Radius from $NICER$ Data and Implications
  for the Properties of Neutron Star Matter}.
\newblock {\em Astrophys. J. Lett.} {\bf 2019}, {\em 887},~L24,
  \href{http://xxx.lanl.gov/abs/1912.05705}{{\normalfont
  [arXiv:astro-ph.HE/1912.05705]}}.
\newblock {\url{https://doi.org/10.3847/2041-8213/ab50c5}}.

\bibitem[Riley et~al.(2021)]{Riley:2021pdl}
Riley, T.E.;  et~al.
\newblock {A NICER View of the Massive Pulsar PSR J0740+6620 Informed by Radio
  Timing and XMM-Newton Spectroscopy}.
\newblock {\em Astrophys. J. Lett.} {\bf 2021}, {\em 918},~L27,
  \href{http://xxx.lanl.gov/abs/2105.06980}{{\normalfont
  [arXiv:astro-ph.HE/2105.06980]}}.
\newblock {\url{https://doi.org/10.3847/2041-8213/ac0a81}}.

\bibitem[Watts(2019)]{Watts:2019lbs}
Watts, A.L.
\newblock {Constraining the neutron star equation of state using Pulse Profile
  Modeling}.
\newblock {\em AIP Conf. Proc.} {\bf 2019}, {\em 2127},~020008,
  \href{http://xxx.lanl.gov/abs/1904.07012}{{\normalfont
  [arXiv:astro-ph.HE/1904.07012]}}.
\newblock {\url{https://doi.org/10.1063/1.5117798}}.

\bibitem[Miao et~al.(2022)Miao, Zhu, Li, and Huang]{Miao:2022rqj}
Miao, Z.; Zhu, Y.; Li, A.; Huang, F.
\newblock {Dark Matter Admixed Neutron Star Properties in the Light of X-Ray
  Pulse Profile Observations}.
\newblock {\em Astrophys. J.} {\bf 2022}, {\em 936},~69,
  \href{http://xxx.lanl.gov/abs/2204.05560}{{\normalfont
  [arXiv:astro-ph.HE/2204.05560]}}.
\newblock {\url{https://doi.org/10.3847/1538-4357/ac8544}}.

\bibitem[Shakeri and Hajkarim(2023)]{Shakeri:2022usk}
Shakeri, S.; Hajkarim, F.
\newblock {Probing axions via light circular polarization and event horizon
  telescope}.
\newblock {\em JCAP} {\bf 2023}, {\em 04},~017,
  \href{http://xxx.lanl.gov/abs/2209.13572}{{\normalfont
  [arXiv:hep-ph/2209.13572]}}.
\newblock {\url{https://doi.org/10.1088/1475-7516/2023/04/017}}.

\bibitem[Chavanis(2023)]{Chavanis:2022fvh}
Chavanis, P.H.
\newblock {Maximum mass of relativistic self-gravitating Bose-Einstein
  condensates with repulsive or attractive |\ensuremath{\varphi}|4
  self-interaction}.
\newblock {\em Phys. Rev. D} {\bf 2023}, {\em 107},~103503,
  \href{http://xxx.lanl.gov/abs/2211.13237}{{\normalfont
  [arXiv:gr-qc/2211.13237]}}.
\newblock {\url{https://doi.org/10.1103/PhysRevD.107.103503}}.

\bibitem[Colpi et~al.(1986)Colpi, Shapiro, and Wasserman]{Colpi:1986ye}
Colpi, M.; Shapiro, S.; Wasserman, I.
\newblock {Boson Stars: Gravitational Equilibria of Selfinteracting Scalar
  Fields}.
\newblock {\em Phys. Rev. Lett.} {\bf 1986}, {\em 57},~2485--2488.
\newblock {\url{https://doi.org/10.1103/PhysRevLett.57.2485}}.

\bibitem[Visinelli(2021)]{Visinelli:2021uve}
Visinelli, L.
\newblock {Boson Stars and Oscillatons: A Review} {\bf 2021}.
\newblock  \href{http://xxx.lanl.gov/abs/2109.05481}{{\normalfont
  [arXiv:gr-qc/2109.05481]}}.

\bibitem[Liebling and Palenzuela(2017)]{Liebling:2012fv}
Liebling, S.L.; Palenzuela, C.
\newblock {Dynamical Boson Stars}.
\newblock {\em Living Rev. Rel.} {\bf 2017}, {\em 20},~5,
  \href{http://xxx.lanl.gov/abs/1202.5809}{{\normalfont
  [arXiv:gr-qc/1202.5809]}}.
\newblock {\url{https://doi.org/10.12942/lrr-2012-6}}.

\bibitem[Typel et~al.(2010)Typel, Ropke, Klahn, Blaschke, and
  Wolter]{Typel:2009sy}
Typel, S.; Ropke, G.; Klahn, T.; Blaschke, D.; Wolter, H.H.
\newblock {Composition and thermodynamics of nuclear matter with light
  clusters}.
\newblock {\em Phys. Rev. C} {\bf 2010}, {\em 81},~015803,
  \href{http://xxx.lanl.gov/abs/0908.2344}{{\normalfont
  [arXiv:nucl-th/0908.2344]}}.
\newblock {\url{https://doi.org/10.1103/PhysRevC.81.015803}}.

\bibitem[Del~Popolo et~al.(2020)Del~Popolo, Deliyergiyev, and
  Le~Delliou]{DelPopolo:2020pzh}
Del~Popolo, A.; Deliyergiyev, M.; Le~Delliou, M.
\newblock {Solution to the hyperon puzzle using dark matter}.
\newblock {\em Phys. Dark Univ.} {\bf 2020}, {\em 30},~100622,
  \href{http://xxx.lanl.gov/abs/2011.00984}{{\normalfont
  [arXiv:gr-qc/2011.00984]}}.
\newblock {\url{https://doi.org/10.1016/j.dark.2020.100622}}.

\bibitem[Ferreira and Fraga(2023)]{Ferreira:2022fjo}
Ferreira, O.; Fraga, E.S.
\newblock {Strange magnetars admixed with fermionic dark matter}.
\newblock {\em JCAP} {\bf 2023}, {\em 04},~012,
  \href{http://xxx.lanl.gov/abs/2209.10959}{{\normalfont
  [arXiv:hep-ph/2209.10959]}}.
\newblock {\url{https://doi.org/10.1088/1475-7516/2023/04/012}}.

\bibitem[Lenzi et~al.(2023)Lenzi, Dutra, Louren\c{c}o, Lopes, and
  Menezes]{Lenzi:2022ypb}
Lenzi, C.H.; Dutra, M.; Louren\c{c}o, O.; Lopes, L.L.; Menezes, D.P.
\newblock {Dark matter effects on hybrid star properties}.
\newblock {\em Eur. Phys. J. C} {\bf 2023}, {\em 83},~266,
  \href{http://xxx.lanl.gov/abs/2212.12615}{{\normalfont
  [arXiv:hep-ph/2212.12615]}}.
\newblock {\url{https://doi.org/10.1140/epjc/s10052-023-11416-y}}.

\bibitem[Yang et~al.(2023)Yang, Pi, Zheng, and Weber]{Yang:2023haz}
Yang, S.H.; Pi, C.M.; Zheng, X.P.; Weber, F.
\newblock {Confronting Strange Stars with Compact-Star Observations and New
  Physics}.
\newblock {\em Universe} {\bf 2023}, {\em 9},~202,
  \href{http://xxx.lanl.gov/abs/2304.09614}{{\normalfont
  [arXiv:astro-ph.HE/2304.09614]}}.
\newblock {\url{https://doi.org/10.3390/universe9050202}}.

\bibitem[Lopes and Das(2023)]{Lopes:2023uxi}
Lopes, L.L.; Das, H.C.
\newblock {Strange stars within bosonic and fermionic admixed dark matter}.
\newblock {\em JCAP} {\bf 2023}, {\em 05},~034,
  \href{http://xxx.lanl.gov/abs/2301.00567}{{\normalfont
  [arXiv:astro-ph.HE/2301.00567]}}.
\newblock {\url{https://doi.org/10.1088/1475-7516/2023/05/034}}.

\bibitem[Lopes et~al.(2022)Lopes, Farias, Dexheimer, Bandyopadhyay, and
  O.~Ramos]{Lopes:2022efy}
Lopes, B.S.; Farias, R.L.S.; Dexheimer, V.; Bandyopadhyay, A.; O.~Ramos, R.
\newblock {Axion effects in the stability of hybrid stars}.
\newblock {\em Phys. Rev. D} {\bf 2022}, {\em 106},~L121301,
  \href{http://xxx.lanl.gov/abs/2206.01631}{{\normalfont
  [arXiv:hep-ph/2206.01631]}}.
\newblock {\url{https://doi.org/10.1103/PhysRevD.106.L121301}}.

\bibitem[Sen and Guha(2022)]{Sen:2022pfr}
Sen, D.; Guha, A.
\newblock {Vector dark boson mediated feeble interaction between fermionic dark
  matter and strange quark matter in quark stars}.
\newblock {\em Mon. Not. Roy. Astron. Soc.} {\bf 2022}, {\em 517},~518--525,
  \href{http://xxx.lanl.gov/abs/2209.09021}{{\normalfont
  [arXiv:hep-ph/2209.09021]}}.
\newblock {\url{https://doi.org/10.1093/mnras/stac2675}}.

\bibitem[Jim\'enez and Fraga(2022)]{Jimenez:2021nmr}
Jim\'enez, J.C.; Fraga, E.S.
\newblock {Radial Oscillations of Quark Stars Admixed with Dark Matter}.
\newblock {\em Universe} {\bf 2022}, {\em 8},~34,
  \href{http://xxx.lanl.gov/abs/2111.00091}{{\normalfont
  [arXiv:hep-ph/2111.00091]}}.
\newblock {\url{https://doi.org/10.3390/universe8010034}}.

\bibitem[Miller et~al.(2021)]{Miller:2021qha}
Miller, M.C.;  et~al.
\newblock {The Radius of PSR J0740+6620 from NICER and XMM-Newton Data}.
\newblock {\em Astrophys. J. Lett.} {\bf 2021}, {\em 918},~L28,
  \href{http://xxx.lanl.gov/abs/2105.06979}{{\normalfont
  [arXiv:astro-ph.HE/2105.06979]}}.
\newblock {\url{https://doi.org/10.3847/2041-8213/ac089b}}.

\bibitem[Romani et~al.(2022)Romani, Kandel, Filippenko, Brink, and
  Zheng]{Romani:2022jhd}
Romani, R.W.; Kandel, D.; Filippenko, A.V.; Brink, T.G.; Zheng, W.
\newblock {PSR J0952\ensuremath{-}0607: The Fastest and Heaviest Known Galactic
  Neutron Star}.
\newblock {\em Astrophys. J. Lett.} {\bf 2022}, {\em 934},~L18,
  \href{http://xxx.lanl.gov/abs/2207.05124}{{\normalfont
  [arXiv:astro-ph.HE/2207.05124]}}.
\newblock {\url{https://doi.org/10.3847/2041-8213/ac8007}}.

\bibitem[Cromartie et~al.(2019)]{NANOGrav:2019jur}
Cromartie, H.T.;  et~al.
\newblock {Relativistic Shapiro delay measurements of an extremely massive
  millisecond pulsar}.
\newblock {\em Nature Astron.} {\bf 2019}, {\em 4},~72--76,
  \href{http://xxx.lanl.gov/abs/1904.06759}{{\normalfont
  [arXiv:astro-ph.HE/1904.06759]}}.
\newblock {\url{https://doi.org/10.1038/s41550-019-0880-2}}.

\bibitem[Riley et~al.(2019)]{Riley:2019yda}
Riley, T.E.;  et~al.
\newblock {A $NICER$ View of PSR J0030+0451: Millisecond Pulsar Parameter
  Estimation}.
\newblock {\em Astrophys. J. Lett.} {\bf 2019}, {\em 887},~L21,
  \href{http://xxx.lanl.gov/abs/1912.05702}{{\normalfont
  [arXiv:astro-ph.HE/1912.05702]}}.
\newblock {\url{https://doi.org/10.3847/2041-8213/ab481c}}.

\bibitem[Dietrich et~al.(2020)Dietrich, Coughlin, Pang, Bulla, Heinzel, Issa,
  Tews, and Antier]{Dietrich:2020efo}
Dietrich, T.; Coughlin, M.W.; Pang, P.T.H.; Bulla, M.; Heinzel, J.; Issa, L.;
  Tews, I.; Antier, S.
\newblock {Multimessenger constraints on the neutron-star equation of state and
  the Hubble constant}.
\newblock {\em Science} {\bf 2020}, {\em 370},~1450--1453,
  \href{http://xxx.lanl.gov/abs/2002.11355}{{\normalfont
  [arXiv:astro-ph.HE/2002.11355]}}.
\newblock {\url{https://doi.org/10.1126/science.abb4317}}.

\bibitem[Abbott et~al.(2017)Abbott, Abbott, Abbott, Acernese, Ackley, Adams,
  Adams, Addesso, Adhikari, Adya, and et~al.]{Abbott_2017}
Abbott, B.; Abbott, R.; Abbott, T.; Acernese, F.; Ackley, K.; Adams, C.; Adams,
  T.; Addesso, P.; Adhikari, R.; Adya, V.;  et~al.
\newblock GW170817: Observation of Gravitational Waves from a Binary Neutron
  Star Inspiral.
\newblock {\em Physical Review Letters} {\bf 2017}, {\em 119}.
\newblock {\url{https://doi.org/10.1103/physrevlett.119.161101}}.

\bibitem[Hinderer et~al.(2010)Hinderer, Lackey, Lang, and
  Read]{Hinderer:2009ca}
Hinderer, T.; Lackey, B.D.; Lang, R.N.; Read, J.S.
\newblock {Tidal deformability of neutron stars with realistic equations of
  state and their gravitational wave signatures in binary inspiral}.
\newblock {\em Phys. Rev. D} {\bf 2010}, {\em 81},~123016,
  \href{http://xxx.lanl.gov/abs/0911.3535}{{\normalfont
  [arXiv:astro-ph.HE/0911.3535]}}.
\newblock {\url{https://doi.org/10.1103/PhysRevD.81.123016}}.

\bibitem[Abbott et~al.(2018)]{Abbott:2018exr}
Abbott, B.P.;  et~al.
\newblock {GW170817: Measurements of neutron star radii and equation of state}.
\newblock {\em Phys. Rev. Lett.} {\bf 2018}, {\em 121},~161101,
  \href{http://xxx.lanl.gov/abs/1805.11581}{{\normalfont
  [arXiv:gr-qc/1805.11581]}}.
\newblock {\url{https://doi.org/10.1103/PhysRevLett.121.161101}}.

\bibitem[Thakur et~al.(2024)Thakur, Malik, and Jha]{Thakur:2024mxs}
Thakur, P.; Malik, T.; Jha, T.K.
\newblock {Towards Uncovering Dark Matter Effects on Neutron Star Properties: A
  Machine Learning Approach}.
\newblock {\em Particles} {\bf 2024}, {\em 7},~80--95,
  \href{http://xxx.lanl.gov/abs/2401.07773}{{\normalfont
  [arXiv:hep-ph/2401.07773]}}.
\newblock {\url{https://doi.org/10.3390/particles7010005}}.

\bibitem[Xiang et~al.(2014)Xiang, Jiang, Zhang, and Yang]{Xiang:2013xwa}
Xiang, Q.F.; Jiang, W.Z.; Zhang, D.R.; Yang, R.Y.
\newblock {Effects of fermionic dark matter on properties of neutron stars}.
\newblock {\em Phys. Rev. C} {\bf 2014}, {\em 89},~025803,
  \href{http://xxx.lanl.gov/abs/1305.7354}{{\normalfont
  [arXiv:astro-ph.SR/1305.7354]}}.
\newblock {\url{https://doi.org/10.1103/PhysRevC.89.025803}}.

\bibitem[Oertel et~al.(2017)Oertel, Hempel, Kl\"ahn, and Typel]{Oertel:2016bki}
Oertel, M.; Hempel, M.; Kl\"ahn, T.; Typel, S.
\newblock {Equations of state for supernovae and compact stars}.
\newblock {\em Rev. Mod. Phys.} {\bf 2017}, {\em 89},~015007,
  \href{http://xxx.lanl.gov/abs/1610.03361}{{\normalfont
  [arXiv:astro-ph.HE/1610.03361]}}.
\newblock {\url{https://doi.org/10.1103/RevModPhys.89.015007}}.

\bibitem[Hinderer(2008)]{Hinderer:2007mb}
Hinderer, T.
\newblock {Tidal Love numbers of neutron stars}.
\newblock {\em Astrophys. J.} {\bf 2008}, {\em 677},~1216--1220,
  \href{http://xxx.lanl.gov/abs/0711.2420}{{\normalfont
  [arXiv:astro-ph/0711.2420]}}.
\newblock {\url{https://doi.org/10.1086/533487}}.

\end{thebibliography}

\PublishersNote{}
\end{adjustwidth}
\end{document}